\documentclass[twocolumn]{revtex4-1}

\usepackage{amsmath,amssymb}
\usepackage{graphicx}
\usepackage{dcolumn}% Align table columns on decimal point
\usepackage{bm}% bold math
\usepackage{multirow}
\usepackage{braket}

\def \n {\nonumber\\}

\usepackage{tikz}
\usetikzlibrary{shapes,arrows}

\begin{document}
\title{Many-body theory of optical absorption in doped two-dimensional semiconductors}
\author{Yao-Wen Chang}
\email{yaowen920@gmail.com}
\author{David R. Reichman}
\email{drr2103@columbia.edu}
\affiliation{Department of Chemistry, Columbia University, New York, New York 10027, USA}
\date{\today}
\begin{abstract}
  In this article, we use a many-body approach to study the absorption spectra of
  electron-doped two-dimensional semiconductors. Optical absorption is modeled by a
  many-body scattering Hamiltonian which describes an exciton immersed in a Fermi sea. The
  interaction between electron and exciton is approximated by an effective scattering
  potential, and optical spectra are calculated by solving for the exciton Green's
  function. From this approach, a trion state can be assigned as a bound state of an
  electron-exciton scattering process, and the doping-dependent phenomena observed in the
  spectra can be attributed to several many-body effects induced by the interaction with
  the Fermi sea. While the many-body scattering Hamiltonian can not solved exactly, we
  reduce the problem to two limiting solvable situations. The first approach approximates
  the full many-body problem by a simple scattering process between the electron and the
  exciton, with a self-energy obtained by solving a Bethe-Salpeter equation (BSE). An
  alternate approach assumes an infinite mass for the exciton, such that the many-body
  scattering Hamiltonian reduces to a Mahan-Nozi\'{e}res-De Dominicis (MND) model. The
  exciton Green's function can then be solved numerically exactly by a determinantal
  formulation, and the optical spectra show signatures of the Fermi-edge singularity at
  high doping densities. The full doping dependence and temperature dependence of the
  exciton and trion lineshapes are simulated via these two approximate approaches, with
  the results compared to each other and to experimental expectations.
\end{abstract}

\maketitle

\section{Introduction}

A trion is a three-particle bound state which is composed of two electrons and one hole or
two holes and one electron\cite{kheng1993observation}. It can also be viewed as a
negatively or positively charged exciton, which is generated by optical absorption in
doped semiconductors or nanostructures. In the absorption or photoluminescence spectra of
such systems at low doping density, the trion peak is observed on the low energy side of
the exciton peak. The energy difference between the trion and exciton peaks in the zero
doping-density limit is called the trion binding energy (denoted as $\Delta_{\text{T}}$).
For semiconducting quantum wells, this binding energy is a few
meV\cite{kheng1993observation, huard2000bound, astakhov2000oscillator,
ciulin2000radiative}. The trion binding energy rises to about 30 meV for monolayer
transition metal dichalcogenides (TMDCs)\cite{mak2013tightly, berkelbach2013theory,
yang2015robust, berkelbach2017optical} and about 100 meV for single-wall carbon
nanotubes\cite{akizuki2014nonlinear, hartleb2015evidence}.

While the nature of trion state in the limit of vanishing doping density is well studied
and well understood, the physical description of the "trion transition" as a function of
doping remains unclear\cite{combescot2003trion, combescot2005many}. A trion transition is
an optical process of trion creation by photon absorption or trion annihilation via
emission of light. In this discussion we consider specifically the optical absorption of
electron-doped materials. Assuming electron doping can be described by altering the Fermi
level in the conduction band, one can use the Fermi energy (denoted as
$\varepsilon_{\text{F}}$) to estimate the scale of the electron doping density. According
to various experimental measurements\cite{kheng1993observation, huard2000bound,
astakhov2000oscillator, ciulin2000radiative, mak2013tightly, yang2015robust}, the
lineshapes and intensities of exciton and trion transitions are notably affected by the
doping density. For example as $\varepsilon_{\text{F}}$ increases, the exciton peak
diminishes and the trion peak amplifies. The exciton peak begins to be depleted as the
Fermi energy exceeds the trion binding energy
($\varepsilon_{\text{F}}\geq\Delta_{\text{T}}$). The total area of the combination of the
two peaks in the absorption spectrum is relatively insensitive to the doping density
variation. This behavior is connected to a sum rule that conserves the total oscillator
strength. In addition, the energy splitting between the exciton peak and trion peak grows
with increasing doping density, and is roughly given by
$E_{\text{split}}\simeq\Delta_{\text{T}}+\varepsilon_{\text{F}}$. Some experiments also
show that the exciton linewidth increases proportionally to the Fermi energy, while the
trion linewidth is relatively insensitive to it\cite{astakhov2000oscillator}. Finally, in
the high doping density regime, the trion and exciton lineshapes are susceptible to
temperature, and may become strongly asymmetric\cite{huard2000bound,
astakhov2000oscillator}. This latter feature implies the possible existence of a
Fermi-edge singularity, which has been shown to exist in doped two-dimensional
semiconductors\cite{ruckenstein1987many, hawrylak1991optical, brown1996evolution}. All of
these doping-dependent phenomena suggest the importance of many-body effects on the trion
transition.

Various theories have been proposed to study the trion transition and simulate the
doping-dependent optical spectra in semiconductors\cite{combescot2003trion,
stebe1998optical, esser2000photoluminescence, bronold2000absorption, esser2001theory,
suris2001excitons, ossau2012optical, shiau2012trion, baeten2015many, efimkin2017many}, but
many questions remain. A common picture of the trion transition is that of an
electron-hole-pair generated in the presence of a background doping electron, from which
the bound trion state forms. The transition amplitude connecting the exciton, the
background electron and the trion state can be calculated by Fermi's golden
rule\cite{combescot2003trion, stebe1998optical, esser2000photoluminescence,
shiau2012trion}. This method is useful in the extremely low doping-density limit
($\varepsilon_{\text{F}}\ll\Delta_{\text{T}}$), but it fails to explain doping-dependent
phenomena for higher doping densities.

An improved approach proposed by Bronold\cite{bronold2000absorption, ossau2012optical}
suggests a dynamical theory to explain the oscillator-strength competition between the
exciton and trion transitions. Accordingly, the photogenerated valence hole scatters and
transfers momentum to excite a Fermi sea electron-hole pair in the conduction band. The
excited electron interacts with the photogenerated electron-hole pair to form a bound
trion state. Along with the hole in the Fermi sea, the trion transition is interpreted as
a dynamical trion-hole generation process. Bronold employed an exciton Green's function
formalism to simulate the absorption lineshape, and included the dynamical trion-hole
generation by means of diagrammatic self-energy corrections. Within this formalism, the
doping-dependent optical spectrum can be simulated in low doping-density regime
($\varepsilon_{\text{F}}<\Delta_{\text{T}}$), and the sum rule and oscillator strength
transfer can be elucidated well. In addition to this study, Esser \textit{et al.} also
proposed a similar dynamical theory based on the density-matrix
approach\cite{esser2001theory}. They applied their theory to the optical spectra of
one-dimensional nanostructures and found a good correspondence with expectations from
experimental work. However, a detailed discussion of the trion and exciton lineshapes with
respect to different doping densities and temperatures in two dimensions has not been
carried out within these theoretical frameworks.

Other theories exist which describe the dynamical process of trion-hole generation. One
such theory is the T-matrix model proposed by Suris \textit{et
al.}\cite{suris2001excitons, ossau2012optical}. Based on this approach, the trion binding
energy can be obtained by solving a two-particle Lippmann-Schwinger equation which
describes electron-exciton scattering. By assuming that only s-wave scattering is
important and parametrizing the T-matrix phenomenologically, the authors found reasonable
doping-dependent lineshape variation without concern for the details of the scattering
potential. Recently, Efimkin and MacDonald extended the T-matrix model to create a
Fermi-polaron theory of trion transition\cite{efimkin2017many}. The Fermi-polaron picture
is normally employed in the study of the quasi-particle properties of impurity atoms
immersed in, and strongly coupled to, an ultracold Fermi gas\cite{klawunn2011fermi,
schmidt2012fermi, parish2013highly, massignan2014polarons}. In the present context, the
impurity atom is replaced by an exciton and the Fermi gas is represented by the electron
gas, such that the quasi-particle is given the name "exciton-polaron." In this framework,
the trion transition can be interpreted as the lower-energy attractive exciton-polaron
branch. A derivable scattering model has been built by approximating the electron-exciton
scattering potential by a contact interaction. The trion binding energy
($\Delta_{\text{T}}$) can then be calculated and the optical spectrum in full
doping-density regime can also be obtained. However, the trion binding energy appears to
be overestimated and is found to be dependent on an ultraviolet cut-off energy. Since the
cut-off dependence is abruptly distinguishable with a trion model described by the
three-particle Schr\"{o}dinger-equation (presumably valid in the limit of vanishing doping
density), the difference between the exciton-polaron state and the standard view of a
trion state requires further investigation.

Based on concepts related to the Fermi-polaron approach, the Fermi-edge singularity
associated with optical transitions in doped materials in the large exciton mass limit can
also be studied. In particular, the Mahan-Nozi\'{e}res-De Dominicis (MND) theory provides
a framework to study the doping dependence of optical lineshapes in a dense electron
gas\cite{mahan1967excitons, ruckenstein1987many, nozieres1969singularities,
gavoret1969optical, combescot1971infrared, mahan1980final, von1982dynamical,
hawrylak1991optical, mahan2013many, pimenov2017fermi}. Traditionally, the MND model is
used to study an infinite-mass hole immersed in a Fermi sea in conjunction with an
electron-hole scattering potential to explain the origin of edge singularity behavior. In
a recent work, Baeten and Wouters applied an electron-exciton-scattering version of MND
theory to the study of trion-polaritons\cite{baeten2015many}. A long-range attractive
Yukawa potential was used to simulate the electron-exciton interaction, and a trion
transition can be obtained by numerically solving the model. The calculated
doping-dependent optical spectra show some features coincident with behaviors observed in
the optical spectra of TMDCs. However, the use of a Yukawa form of the scattering
potential results in an overestimation of the trion binding energy and a
doping-independent exciton linewidth. A better choice of model potential to describe
electron-exciton scattering would be helpful to remedy these issues for applications in
two-dimensional semiconductors.

The goal of the present work is to unify our understanding of the doping-dependent optical
spectra of two-dimensional semiconductors via a systematic exploration of these related
approaches. An appropriate electron-exciton scattering potential is derived, and a
many-body scattering Hamiltonian which describes an exciton immersed in Fermi sea
interacting with electrons through this potential is written down. While the many-body
Hamiltonian can not be solved exactly, we reduce the description to two limiting
situations which correspond to a particular type of electron-exciton scattering problem
and the MND problem. The electron-exciton scattering problem can be solved by a
Bethe-Salpeter equation (BSE), whose eigenspectrum can be included into the self-energy in
a Green's function formalism. We show that this approach, which we will call the BSE
formalism in the following discussions, is conceptually equivalent to Bronold's dynamical
theory of trion-hole generation, Suris' T-matrix model, and Efimkin-MacDonald's
Fermi-polaron theory. The doping-dependent exciton and trion lineshapes can be studied
analytically in low doping-density regime, and extended to higher doping densities
numerically. In the other limiting situation of infinite exciton mass the MND model is
numerically solved and the optical spectrum is obtained by propagating the Green's
function in the time domain. The doping-dependent and temperature-dependent optical
spectra of two-dimensional materials can then be numerically studied and compared to that
of the BSE approach and to experimental expectations.

This article is organized as follows. In Sec.~\ref{sec:wavefunction}, we give a heuristic
review of the theories of optical transitions in doped semiconductors by writing down the
wavefunctions associated with the optically-generated quasi-particles. The concepts and
relationship between the exciton state, trion state, trion transition, electron-exciton
scattering, dynamical trion-hole generation, Fermi-polaron and Fermi-edge singularity are
introduced in a second-quantization language. Building on these concepts, the derivation
of the electron-exciton scattering potential becomes manifest. In Sec.~\ref{sec:theory},
the electron-exciton scattering potential is derived and the many-body scattering
Hamiltonian is written down. Approximate methods based on the BSE formalism and the MND
theory to solve the Hamiltonian are introduced. The connection between the BSE formalism
and Efimkin-MacDonald's Fermi-polaron theory is also discussed. By using the self-energy
from the BSE formalism, the exciton and trion transitions are analyzed in
Sec.~\ref{sec:transitions}. Numerical calculations within both the BSE formalism and the
MND theory are provided in Sec.~\ref{sec:numerical} to study the exciton and trion
lineshapes. The optical energy renormalization and Pauli-blocking effects due to
electron-doping in two-dimensional materials are also discussed.  Finally, our conclusions
are given in Sec.~\ref{sec:conclusion}.

\section{Heuristic wavefunction theory\label{sec:wavefunction}}

In the case of weak light-matter interaction, the optical absorption of solid-state
materials can be realized as a dynamical conduction process. The creation of an exciton is
associated with the electron-hole-pair generation process, whereby the electron and hole
become bound due to the attractive Coulomb interaction between them. The
frequency-dependent transition spectral density of this process is proportional to
\begin{eqnarray}
  \mathcal{A}(\omega)\sim\frac{1}{\omega}|\braket{\text{X}_{\mathbf{K}}|\hat{j}|0}|^2
  \delta(\omega-\varepsilon_{\text{X},\mathbf{K}}),
\end{eqnarray}
where $\ket{\text{X}_{\mathbf{K}}}$ is an exciton state, $\ket{0}$ is the ground state,
$\varepsilon_{\text{X},\mathbf{K}}
=\braket{\text{X}_{\mathbf{K}}|\hat{\mathcal{H}}|\text{X}_{\mathbf{K}}}
-\braket{0|\hat{\mathcal{H}}|0}$ is the exciton transition energy, and $\hat{j}$ is the
current operator\cite{mahan2013many}. The ground state can be described as one with the
valence band fully occupied and the conduction band completely empty. The exciton state
can then be written as
$\ket{\text{X}_{\mathbf{K}}}=\hat{X}^{\dagger}_{\mathbf{K}}\ket{0}$, where
\begin{eqnarray}
  \hat{X}^{\dagger}_{\mathbf{K}}
  &=&
  \sum_{\mathbf{p}}\Psi_{\text{X},\mathbf{p}}
  \hat{c}^{\dagger}_{\mathbf{p}+\mathbf{K}}\hat{d}^{\dagger}_{-\mathbf{p}},
\end{eqnarray}
is the exciton creation operator, $\hat{c}^{\dagger}_{\mathbf{p}}$ is the creation
operator of electron in the conduction band with quasi-momentum $\mathbf{p}$,
$\hat{d}^{\dagger}_{\mathbf{p}}$ is the creation operator of hole in the valence band, and
$\Psi_{\text{X},\mathbf{p}}$ is the exciton wavefunction. For a direct band-gap
semiconductor, the electric current operator is written as
\begin{eqnarray}
  \hat{{j}}
  \simeq
  e\sum_{\mathbf{k}}\left(\mathcal{P}_{\mathbf{k}\text{cv}}
  \hat{c}^{\dagger}_{\mathbf{k}}\hat{d}^{\dagger}_{\mathbf{k}}
  +\mathcal{P}^*_{\mathbf{k}\text{cv}}\hat{d}_{\mathbf{k}}\hat{c}_{\mathbf{k}}\right),
\end{eqnarray}
where $\mathcal{P}_{\mathbf{k}\text{cv}}$ is the momentum matrix element. Therefore, a
vertical transition selection rule $\braket{\text{X}_{\mathbf{0}}|\hat{j}|0}\neq{0}$ can
be found.

On the other hand, the trion transition is more complex. The trion state can be written as
$\ket{\text{T}_{\mathbf{Q}}}=\hat{T}^{\dagger}_{\mathbf{Q}}\ket{0}$, where the trion
creation operator is
\begin{eqnarray}
  \hat{T}^{\dagger}_{\mathbf{Q}}
  &=&
  \sum_{\mathbf{p}\mathbf{q}}\Psi_{\text{T},\mathbf{p}\mathbf{q}}
  \hat{c}^{\dagger}_{\mathbf{p}+\frac{m_{\text{e}}}{M_{\text{T}}}\mathbf{Q}}
  \hat{c}^{\dagger}_{\mathbf{q}+\frac{m_{\text{e}}}{M_{\text{T}}}\mathbf{Q}}
  \hat{d}^{\dagger}_{-\mathbf{p}-\mathbf{q}+\frac{m_{\text{h}}}{M_{\text{T}}}\mathbf{Q}},
\end{eqnarray}
with $\Psi_{\text{T},\mathbf{p}\mathbf{q}}$ the trion wavefunction, $m_{\text{e}}$ the
electron mass, $m_{\text{e}}$ the hole mass, and $M_{\text{T}}=2m_{\text{e}}+m_{\text{h}}$
the trion mass. With these assumptions, it is found that the transition amplitude
$\braket{\text{T}_{\mathbf{Q}}|\hat{j}|0}$ is zero. A commonly used modification is to
employ a different initial state in calculating the transition rate. Assuming that the
initial state is $\ket{\text{e}_{\mathbf{Q}}}=\hat{c}^{\dagger}_{\mathbf{Q}}\ket{0}$, such
that the transition amplitude
$\braket{\text{T}_{\mathbf{Q}}|\hat{j}|\text{e}_{\mathbf{Q}}}$ is non-zero, then the
transition spectral density based on Fermi's golden rule is\cite{stebe1998optical,
esser2000photoluminescence, shiau2012trion}
\begin{eqnarray}
  \mathcal{A}(\omega)
  &\sim&
  \frac{1}{\omega}\left(1-\sigma\sum\nolimits_{\mathbf{Q}}n_{\mathbf{Q}}\right)
  |\braket{\text{X}_{\mathbf{0}}|\hat{j}|0}|^2
  \delta\left(\omega-\varepsilon_{\text{X}}\right)\n
  &&+\frac{\sigma}{\omega}
  \sum_{\mathbf{Q}}n_{\mathbf{Q}}|\braket{\text{T}_{\mathbf{Q}}|\hat{j}|
  \text{e}_{\mathbf{Q}}}|^2\delta\left(\omega-\varepsilon_{\text{T},\mathbf{Q}}
  +\varepsilon_{\text{e},\mathbf{Q}}\right),\n
  \label{fermi_golden_rule}
\end{eqnarray}
where $\sigma$ is an adjustable parameter related to the photon-electron scattering
cross-section, $n_{\mathbf{Q}}$ is the electron density distribution of the
$\ket{\text{e}_{\mathbf{Q}}}$ state, $\varepsilon_{\text{T},\mathbf{Q}}
=\braket{\text{T}_{\mathbf{Q}}|\hat{\mathcal{H}}|\text{T}_{\mathbf{Q}}}
-\braket{0|\hat{\mathcal{H}}|0}$ is the trion transition energy, and
$\varepsilon_{\text{e},\mathbf{Q}}=\braket{\text{e}_{\mathbf{Q}}|\hat{\mathcal{H}}|
\text{e}_{\mathbf{Q}}}-\braket{0|\hat{\mathcal{H}}|0}$ is the electron quasi-particle
energy. Although this is an \textit{ad hoc} description, the above formula is quite useful
and physical in the extreme low doping-density limit
($\varepsilon_{\text{F}}\ll\Delta_{\text{T}}$). Particularly if we assume
$\braket{\text{X}_{\mathbf{0}}|\hat{j}|0}
\simeq\braket{\text{T}_{\mathbf{Q}}|\hat{j}|\text{e}_{\mathbf{Q}}}$, integration over the
rate constant $\int\text{d}\omega\;\mathcal{A}(\omega)$ is roughly invariant to doping
density, implying that the lowest-order sum rule is fulfilled. However, with higher doping
density and thus $\sum_{\mathbf{Q}}n_{\mathbf{Q}}\gtrsim 1/\sigma$, the transition rate
constant becomes negative and the formula becomes unphysical.

Another description of the trion transition is given by Fermi-polaron
approach\cite{efimkin2017many}. Based on this framework, the ground state is described by
the Fermi sea formed by the electron gas that resides in the conduction band, in addition
to the fully occupied valence band. When an exciton is excited, the Coulomb interaction
between the exciton and the Fermi sea induces electron-hole polarization near the Fermi
surface. The Fermi-polaron state $\ket{\text{P}}$ can be written as\cite{parish2013highly}
\begin{eqnarray}
  \ket{\text{P}}
  &=&
  \Phi_{0}\hat{X}^{\dagger}_{\mathbf{0}}\ket{\text{FS}}
  +\sum_{\mathbf{q},\mathbf{K}}\Phi_{\mathbf{K}}\hat{X}^{\dagger}_{-\mathbf{K}}
  \hat{c}^{\dagger}_{\mathbf{q}+\mathbf{K}}\hat{c}_{\mathbf{q}}\ket{\text{FS}},
  \label{fermi_polaron}
\end{eqnarray}
where $\Phi_{0}$, $\Phi_{\mathbf{K}}$ are superposition coefficients, and
$\ket{\text{FS}}$ is the Fermi sea conduction-band plus fully occupied valence-band state.
The coefficients $\Phi_{0}$, $\Phi_{\mathbf{K}}$ can be evaluated by treating
$\ket{\text{P}}$ as variational wavefunction to minimize the variational energy. Since we
are not considering the Coulomb interaction among the electrons in the Fermi sea, we can
assume the Fermi sea is composed of independent electrons and then replace
$\ket{\text{FS}}$ by $\hat{c}^{\dagger}_{\mathbf{Q}}\ket{\tilde{\text{FS}}}$, where
$\ket{\tilde{\text{FS}}}=\hat{c}_{\mathbf{Q}}\ket{\text{FS}}$. Therefore, the
Fermi-polaron state becomes
\begin{eqnarray}
  \ket{\text{P}}
  &=&
  \Phi_{0}\hat{X}^{\dagger}_{\mathbf{0}}c^{\dagger}_{\mathbf{Q}}\ket{\tilde{\text{FS}}}
  +\sum_{\mathbf{K}}\Phi_{\mathbf{K}}\hat{X}^{\dagger}_{-\mathbf{K}}
  \hat{c}^{\dagger}_{\mathbf{Q}+\mathbf{K}}\ket{\tilde{\text{FS}}}.
\end{eqnarray}
If we define the incoming state and outgoing state as
\begin{eqnarray}
  \ket{\text{in}}
  =
  \hat{X}^{\dagger}_{\mathbf{0}}c^{\dagger}_{\mathbf{Q}}\ket{\tilde{\text{FS}}},\hskip2ex
  \ket{\text{out}}
  =
  \hat{X}^{\dagger}_{-\mathbf{K}}\hat{c}^{\dagger}_{\mathbf{Q}+\mathbf{K}}
  \ket{\tilde{\text{FS}}},
  \label{in_n_out}
\end{eqnarray}
then solving for the variational coefficients reduces to an electron-exciton scattering
problem. If there is an attractive interaction between the electron and the exciton, an
extra bound state may exist, which can be interpreted as the trion state.

Although the Fermi-polaron state and the trion state appear to be different, the two
pictures can be connected if the electron-exciton bound state can be related to the trion
state by a linear transformation with the coefficient $\Phi_{\mathbf{K}}$,
\begin{eqnarray}
  \hat{T}^{\dagger}_{\mathbf{Q}}
  &=&
  \sum_{\mathbf{K}}\Phi_{\mathbf{K}}\hat{X}^{\dagger}_{-\mathbf{K}}
  \hat{c}^{\dagger}_{\mathbf{Q}+\mathbf{K}}.
\end{eqnarray}
In this case, the Fermi-polaron state can be rewritten as
\begin{eqnarray}
  \ket{\text{P}}
  &=&
  \Phi_{\mathbf{0}}\hat{X}^{\dagger}_{\mathbf{0}}\ket{\text{FS}}
  +\sum_{\mathbf{Q}}\hat{T}^{\dagger}_{\mathbf{Q}}\hat{c}_{\mathbf{Q}}\ket{\text{FS}}.
\end{eqnarray}
Therefore, Fermi-polaron generation can be seen as the collective excitation of
trion-hole-pair states near the Fermi surface, where the hole is created by annihilating
an electron in the Fermi sea. If the trion-hole interaction energy is small in compared to
the trion binding energy, the excitation will show particle-like features, such that the
process can be interpreted as trion generation. The collective nature of dynamical
trion-hole generation and the trion-hole interaction will however affect the trion and
exciton transition energies and lineshapes.

The problem of the Fermi-edge singularity is naturally connected to that of the
Fermi-polaron, since they both describe an impurity immersed in, and interacting with, the
Fermi sea. In the case under consideration here, the impurity is an exciton. While the
scattering function formalism describes single electron-hole pair excitation near Fermi
surface, the Fermi-edge singularity is an effect caused by multiple electron-hole pair
excitation. Based on the Fermi-polaron state wavefunction, if the exciton mass is much
larger than the electron mass and the electron-density is sufficiently high, the edge
singularity state will include the multiple-excitation terms as
\begin{eqnarray}
  \ket{\text{ES}}
  &=&
  \Phi_{0}\hat{X}^{\dagger}_{\mathbf{0}}\ket{\text{FS}}
  +\sum_{\mathbf{K}}\Phi_{\mathbf{K}}\hat{X}^{\dagger}_{-\mathbf{K}}
  \hat{B}^{\dagger}_{\mathbf{K}}\ket{\text{FS}}\n
  &&+\sum_{\mathbf{K}_1\mathbf{K}_2}\Phi_{\mathbf{K}_1,\mathbf{K}_2}
  \hat{X}^{\dagger}_{-\mathbf{K}_1-\mathbf{K}_2}
  \left(\hat{B}^{\dagger}_{\mathbf{K}_1}\hat{B}^{\dagger}_{\mathbf{K}_2}\right)
  \ket{\text{FS}}+\cdots\n
  &&+\sum_{\mathbf{K}_1\cdots\mathbf{K}_n}
  \Phi_{\mathbf{K}_1\cdots\mathbf{K}_n}
  \hat{X}^{\dagger}_{-\sum^{n}_{\alpha=1}\mathbf{K}_{\alpha}}
  \left(\prod^{n}_{\alpha'=1}
  \hat{B}^{\dagger}_{\mathbf{K}_{\alpha'}}\right)\ket{\text{FS}}\n
  &&+\cdots,
  \label{edge_singularity}
\end{eqnarray}
where $\hat{B}^{\dagger}_{\mathbf{K}}
=\sum_{\mathbf{q}}\hat{c}^{\dagger}_{\mathbf{q}+\mathbf{K}}\hat{c}_{\mathbf{q}}$ is the
Fermi sea electron-hole excitation operator. If the hole and electron are excited within
the energy scale of the Fermi surface, the Fermi sea electron-hole excitation energy will
be close to zero. When the exciton mass is large, the exciton kinetic energy makes a
vanishingly small contribution to the excitation energy. Therefore, the all
multiple-excitation terms can have an excitation energy coincident with the exciton
transition energy, producing a divergence in the oscillator strength close to the exciton
transition energy.

To describe all relevant transitions, the Fermi-polaron state wavefunction in
Eq.~(\ref{fermi_polaron}) or the edge-singularity state wavefunction in
Eq.~(\ref{edge_singularity}) can act as variational wavefunctions, and the wavefunction
coefficients can be treated as variational parameters. If the lowest-energy solution of
the variation problem has a lower energy than the exciton transition energy, the state may
be interpreted as a trion bound state. Other solutions with energies higher than the
exciton transition energy can be interpreted as electron-exciton scattering states.
However, the variational problem is too difficult to solve in the electron-hole basis due
to the large number of degrees of freedom. Approximations are needed in order to reduce
the numerical effort. In the present work, one approximation we consider is to reduce the
electron-hole basis to an electron-exciton basis. The exciton state is presumed to be a
particle state that can not be decomposed, and the exciton transition energy is
parametrized. Via this approximation, the number of degrees of freedom is greatly reduced.
In Sec.~\ref{sec:theory}, we will give the formal theory of this reduction and provide the
methods of solution.

\section{Formal theory\label{sec:theory}}

In this section, we give a formal theory of a trion transition based on a many-body
scattering Hamiltonian, which includes electron and exciton kinetic energies and an
appropriate electron-exciton scattering potential. We assume that, by solving the exciton
Green's function of the many-body scattering Hamiltonian, the trion transition can be
found and the doping dependence of the lineshapes can be explained. In
Sec.~\ref{sub:scattering_potential} and Sec.~\ref{sub:hamiltonian} the electron-exciton
scattering potential is derived and the many-body scattering Hamiltonian is introduced.
However the many-body Hamiltonian is not exactly solvable, and some approximations must be
employed. One avenue to approximation is the reduction of the many-body problem to a
two-particle scattering problem between an electron and an exciton, with the inclusion of
the scattering processes directly into the exciton self-energy. In
Sec.~\ref{sub:scattering_function}, we introduce this method and show that it is
essentially equivalent to Efimkin and MacDonald's Fermi-polaron theory for optical
absorption of a two-dimensional doped semiconductor\cite{efimkin2017many}. While their
Fermi-polaron theory requires a cut-off energy-dependent trion binding energy, an
alternate method is preferred. In Sec.~\ref{sub:BSE}, a BSE formalism that employs
Bronold's dynamical theory of trion-hole generation\cite{bronold2000absorption} is
derived. Finally, we consider another approximate method valid in the limiting situation
where the exciton mass is infinitely large. Here, the many-body scattering Hamiltonian is
reduced to an electron-exciton version of the MND model\cite{baeten2015many}. In
Sec.~\ref{sub:MND}, the MND model and its numerical method of solution are introduced.

\subsection{Scattering potential\label{sub:scattering_potential}}

In order to find the scattering potential between electron and exciton, we consider the
electron-hole Hamiltonian for electron-doped semiconductors\cite{haug2009quantum}
\begin{eqnarray}
    \hat{\mathcal{H}}
    &=&
    \sum_{\mathbf{k}}\left(\varepsilon_{\text{e},\mathbf{k}}
    \hat{c}^{\dagger}_{\mathbf{k}}\hat{c}_{\mathbf{k}}
    +
    \varepsilon_{\text{h},\mathbf{k}}
    \hat{d}^{\dagger}_{\mathbf{k}}\hat{d}_{\mathbf{k}}\right)\n
    &&+
    \sum_{\mathbf{k}\mathbf{k}'\mathbf{q}}\frac{U_{\mathbf{q}}}{2}\Big(
    \hat{c}^{\dagger}_{\mathbf{k}+\mathbf{q}}
    \hat{c}^{\dagger}_{\mathbf{k}'-\mathbf{q}}\hat{c}_{\mathbf{k}'}
    \hat{c}_{\mathbf{k}}
    -2\hat{c}^{\dagger}_{\mathbf{k}+\mathbf{q}}
    \hat{d}^{\dagger}_{-\mathbf{k}'-\mathbf{q}}\hat{d}_{-\mathbf{k}'}
    \hat{c}_{\mathbf{k}}\Big),\n
\end{eqnarray}
where the Hamiltonian contains kinetic energy terms and Coulomb interaction terms,
$\varepsilon_{\text{e},\mathbf{k}}$, $\varepsilon_{\text{h},\mathbf{k}}$ are the kinetic
energies of electron and hole, and $U_{\mathbf{q}}=v_{\mathbf{q}}/{L^\text{D}}$, with
$v_{\mathbf{q}}$ the Coulomb potential and $L^\text{D}$ the dimension of the system. The
scattering transition amplitudes can be calculated by using the electron-exciton basis
states of Eq.~(\ref{in_n_out}). Note that the $\mathbf{Q}$ of all basis states are equal
in order to fulfill momentum conservation. For the diagonal term of the Hamiltonian
matrix, we find
\begin{eqnarray}
  &&\left(\bra{\tilde{\text{FS}}}\hat{c}_{\mathbf{Q}+\mathbf{K}}\hat{X}_{-\mathbf{K}}
  \right)\hat{\mathcal{H}}\left(
  \hat{X}^{\dagger}_{-\mathbf{K}}\hat{c}^{\dagger}_{\mathbf{Q}+\mathbf{K}}
  \ket{\tilde{\text{FS}}}\right)\n
  &\simeq&
  E_{\tilde{\text{FS}}}+\varepsilon_{\text{X},-\mathbf{K}}
  +\varepsilon_{\text{e},\mathbf{Q}+\mathbf{K}},
\end{eqnarray}
where
$E_{\tilde{\text{FS}}}=\bra{\tilde{\text{FS}}}\hat{\mathcal{H}}\ket{\tilde{\text{FS}}}
=\bra{\text{FS}}\hat{\mathcal{H}}\ket{\text{FS}}-\varepsilon_{\text{e},\mathbf{Q}}$ is the
Fermi sea ground state energy and $\varepsilon_{\text{X},\mathbf{K}}$ is the exciton
excitation energy. The nondiagonal terms of the Hamiltonian matrix give the scattering
potential
\begin{eqnarray}
  V_{\mathbf{K}\mathbf{K}',\mathbf{Q}}
  &=&
  \left(\bra{\tilde{\text{FS}}}\hat{c}_{\mathbf{Q}+\mathbf{K}}\hat{X}_{-\mathbf{K}}
  \right)\hat{\mathcal{H}}\left(
  \hat{X}^{\dagger}_{-\mathbf{K}'}\hat{c}^{\dagger}_{\mathbf{Q}+\mathbf{K}'}
  \ket{\tilde{\text{FS}}}\right).\n
  \label{scattering}
\end{eqnarray}
The exciton creation operator is assumed to be
$\hat{X}^{\dagger}_{-\mathbf{K}}=\sum_{\mathbf{p}}\Psi_{\mathbf{p}}
\hat{c}^{\dagger}_{\mathbf{p}}\hat{d}^{\dagger}_{-\mathbf{p}-\mathbf{K}}$, where
$\Psi_{\mathbf{p}}$ is the exciton wavefunction. Since the electron-exciton interaction
only involves the electron and hole degrees of freedom contained in the basis states,
Eq.~(\ref{scattering}) becomes
\begin{eqnarray}
  V_{\mathbf{K}\mathbf{K}',\mathbf{Q}}
  &=&
  \sum_{\mathbf{p}\mathbf{p}'}\Psi^*_{\mathbf{p}}\Psi_{\mathbf{p}'}\n
  &&\times\braket{\hat{c}_{\mathbf{Q}+\mathbf{K}}
  \hat{c}_{\mathbf{p}}\hat{d}_{-\mathbf{p}-\mathbf{K}}
  :\hat{\mathcal{H}}:
  \hat{c}^{\dagger}_{\mathbf{p}'}
  \hat{d}^{\dagger}_{-\mathbf{p}'-\mathbf{K}'}
  \hat{c}^{\dagger}_{\mathbf{Q}+\mathbf{K}'}},\n
\end{eqnarray}
where $:\hat{\mathcal{H}}:$ indicates that the number operators in the Hamiltonian are
already contracted, i.e.,
$\braket{:\hat{d}^{\dagger}\hat{d}:}=\braket{:\hat{c}^{\dagger}\hat{c}:}=0$. The
electron-electron interaction may be expressed
\begin{widetext}
  \begin{eqnarray}
    \braket{\hat{c}_{\mathbf{Q}+\mathbf{K}}
    \hat{d}_{-\mathbf{p}-\mathbf{K}}\hat{c}_{\mathbf{p}}
    :\hat{\mathcal{H}}_{\text{e-e}}:
    \hat{c}^{\dagger}_{\mathbf{p}'}\hat{d}^{\dagger}_{-\mathbf{p}'-\mathbf{K}'}
    \hat{c}^{\dagger}_{\mathbf{Q}+\mathbf{K}'}}
    &=&
    \sum_{\mathbf{k}\mathbf{k}'\mathbf{q}}\frac{U_{\mathbf{q}}}{2}
    \braket{\hat{c}_{\mathbf{Q}+\mathbf{K}}
    \hat{d}_{-\mathbf{p}-\mathbf{K}}\hat{c}_{\mathbf{p}}
    :\hat{c}^{\dagger}_{\mathbf{k}+\mathbf{q}}
    \hat{c}^{\dagger}_{\mathbf{k}'-\mathbf{q}}\hat{c}_{\mathbf{k}'}
    \hat{c}_{\mathbf{k}}:
    \hat{c}^{\dagger}_{\mathbf{p}'}\hat{d}^{\dagger}_{-\mathbf{p}'-\mathbf{K}'}
    \hat{c}^{\dagger}_{\mathbf{Q}+\mathbf{K}'}}\n
    &=&
    \delta_{\mathbf{p}-\mathbf{p}',\mathbf{K}'-\mathbf{K}}
    \big(U_{\mathbf{K}'-\mathbf{K}}-U_{\mathbf{p}-\mathbf{Q}-\mathbf{K}'}\big).
  \end{eqnarray}
  \begin{eqnarray}
    \braket{\hat{c}_{\mathbf{Q}+\mathbf{K}}
    \hat{d}_{-\mathbf{p}-\mathbf{K}}\hat{c}_{\mathbf{p}}
    :\hat{\mathcal{H}}_{\text{e-h}}:
    \hat{c}^{\dagger}_{\mathbf{p}'}\hat{d}^{\dagger}_{-\mathbf{p}'-\mathbf{K}'}
    \hat{c}^{\dagger}_{\mathbf{Q}+\mathbf{K}'}}
    &=&
    -\sum_{\mathbf{k}\mathbf{k}'\mathbf{q}}U_{\mathbf{q}}
    \braket{\hat{c}_{\mathbf{Q}+\mathbf{K}}
    \hat{d}_{-\mathbf{p}-\mathbf{K}}\hat{c}_{\mathbf{p}}
    :\hat{c}^{\dagger}_{\mathbf{k}+\mathbf{q}}\hat{d}^{\dagger}_{-\mathbf{k}'-\mathbf{q}}
    \hat{d}_{-\mathbf{k}'}\hat{c}_{\mathbf{k}}:
    \hat{c}^{\dagger}_{\mathbf{p}'}\hat{d}^{\dagger}_{-\mathbf{p}'-\mathbf{K}'}
    \hat{c}^{\dagger}_{\mathbf{Q}+\mathbf{K}'}}\n
    &=&
    -\big(U_{\mathbf{p}-\mathbf{p}'}-U_{\mathbf{Q}+\mathbf{K}-\mathbf{p}'}
    \delta_{\mathbf{p},\mathbf{Q}+\mathbf{K}'}
    -U_{\mathbf{p}-\mathbf{Q}-\mathbf{K}'}\delta_{\mathbf{Q}+\mathbf{K},\mathbf{p}'}
    +U_{\mathbf{K}-\mathbf{K}'}\delta_{\mathbf{p},\mathbf{p}'}\big).\n
  \end{eqnarray}
\end{widetext}
The scattering potential can thus be written as
\begin{eqnarray}
  V_{\mathbf{K}\mathbf{K}',\mathbf{Q}}
  &=&
  V^{\text{di-ee}}_{\mathbf{K}\mathbf{K}'}
  +V^{\text{ex-ee}}_{\mathbf{K}\mathbf{K}',\mathbf{Q}}
  +V^{\text{di-eh}}_{\mathbf{K}\mathbf{K}'}
  +V^{\text{ex-eh}}_{\mathbf{K}\mathbf{K}',\mathbf{Q}},
\end{eqnarray}
with a direct electron-electron interaction $V^{\text{di-ee}}_{\mathbf{K}\mathbf{K}'}$, an
exchange electron-electron interaction
$V^{\text{ex-ee}}_{\mathbf{K}\mathbf{K}',\mathbf{Q}}$, a direct electron-hole interaction
$V^{\text{di-eh}}_{\mathbf{K}\mathbf{K}'}$, and an exchange electron-hole interaction
$V^{\text{ex-eh}}_{\mathbf{K}\mathbf{K}',\mathbf{Q}}$. The interactions are
\begin{eqnarray}
  V^{\text{di-ee}}_{\mathbf{K}\mathbf{K}'}
  &=&
  \sum_{\mathbf{p}}U^{*}_{\mathbf{K}-\mathbf{K}'}\Psi^*_{\mathbf{p}}
  \Psi_{\mathbf{p}+\mathbf{K}-\mathbf{K}'},
\end{eqnarray}
\begin{eqnarray}
  V^{\text{ex-ee}}_{\mathbf{K}\mathbf{K}',\mathbf{Q}}
  &=&
  -\sum_{\mathbf{p}}U_{\mathbf{p}}\Psi^*_{\mathbf{p}+\mathbf{Q}+\mathbf{K}'}
  \Psi_{\mathbf{p}+\mathbf{Q}+\mathbf{K}},
\end{eqnarray}
\begin{eqnarray}
  V^{\text{di-eh}}_{\mathbf{K}\mathbf{K}'}
  &=&
  -U_{\mathbf{K}-\mathbf{K}'},
\end{eqnarray}
\begin{eqnarray}
  V^{\text{ex-eh}}_{\mathbf{K}\mathbf{K}',\mathbf{Q}}
  &=&
  \sum_{\mathbf{p}}\big(U^{*}_{\mathbf{p}}
  \Psi^*_{\mathbf{Q}+\mathbf{K}'}\Psi_{\mathbf{p}+\mathbf{Q}+\mathbf{K}}\n
  &&+U_{\mathbf{p}}
  \Psi^*_{\mathbf{p}+\mathbf{Q}+\mathbf{K}'}\Psi_{\mathbf{Q}+\mathbf{K}}\big).
\end{eqnarray}
It it important to note that the present derivation does not consider the spin degree of
freedom. Therefore the electron and hole creation operators contained in the trion
creation operator can be seen to have the same spin quantum number. In this case, the
exchange interaction is of opposite sign to that of the direct interaction. However, if
the two electrons which comprise the electron portion of the trion have different spin
quantum numbers, the exchange interaction can be zero or of the same sign as the direct
interaction, and will depend on the overall spin state of the trion. The spin dependence
of the trion transition is beyond the scope of this work, and we will ignore all exchange
interactions in the following discussion. The effect of spin and exchange interactions
will be studied in future work.

Assuming $U_{\mathbf{q}}=U^{*}_{\mathbf{q}}$ and
$\Delta\mathbf{K}=\mathbf{K}-\mathbf{K}'$, the total direct interaction can be written as
\begin{eqnarray}
  V^{\text{di}}_{\mathbf{K}\mathbf{K}'}
  &=&
  -U_{\Delta\mathbf{K}}
  \left(1-\sum_{\mathbf{p}}\Psi^*_{\mathbf{p}}\Psi_{\mathbf{p}+\Delta\mathbf{K}}\right).
\end{eqnarray}
The wavefunction overlap
$\sum_{\mathbf{p}}\Psi^*_{\mathbf{p}}\Psi_{\mathbf{p}+\Delta\mathbf{K}}$ is unity for
$\Delta\mathbf{K}=0$ and tends to zero as $\Delta\mathbf{K}\rightarrow\infty$. Via Taylor
expansion with respect to $|\Delta\mathbf{K}|$, the wavefunction overlap can be expressed
as
\begin{eqnarray}
  \sum_{\mathbf{p}}\Psi^*_{\mathbf{p}}\Psi_{\mathbf{p}+\Delta\mathbf{K}}
  &\simeq&
  \sum_{\mathbf{p}}\Psi^*_{\mathbf{p}}\Psi_{\mathbf{p}}
  -\mathtt{i}|\Delta\mathbf{K}|\sum_{\mathbf{p}}\Psi^*_{\mathbf{p}}
  \left(\mathtt{i}\frac{\partial}{\partial{p}}\right)\Psi_{\mathbf{p}}\n
  &&-\frac{1}{2}|\Delta\mathbf{K}|^2\sum_{\mathbf{p}}\Psi^*_{\mathbf{p}}
  \left(\mathtt{i}\frac{\partial}{\partial{p}}\right)^2\Psi_{\mathbf{p}}+\cdots.\n
\end{eqnarray}
Assuming the exciton wavefunction is given by the nodeless 1s-orbital wavefunction
obtained from the two-dimensional hydrogen atom problem, the wavefunction overlap can be
approximated as
\begin{eqnarray}
  \sum_{\mathbf{p}}\Psi^*_{\mathbf{p}}\Psi_{\mathbf{p}+\Delta\mathbf{K}}
  &\simeq&
  \exp\left(-\frac{1}{2}|\Delta\mathbf{K}|^2\xi^2\right),
\end{eqnarray}
where $\xi^2=\sum_{\mathbf{p}}\Psi^*_{\mathbf{p}}
\left(\mathtt{i}{\partial}/{\partial{p}}\right)^2\Psi_{\mathbf{p}}$, since it can be shown
that $\sum_{\mathbf{p}}\Psi^*_{\mathbf{p}}
\left(\mathtt{i}{\partial}/{\partial{p}}\right)\Psi_{\mathbf{p}}=0$ for a 1s-orbital. The
exponential form of the approximate wavefunction overlap reproduces the correct behavior
as $\Delta\mathbf{K}\rightarrow{0}$. The characteristic length $\xi$ can be rewritten as
\begin{eqnarray}
  \xi^2
  &=&
  \int\text{d}^{\text{D}}\mathbf{r}\;\tilde{\Psi}^{*}(\mathbf{r})|\mathbf{r}|^2
  \tilde{\Psi}(\mathbf{r}),
\end{eqnarray}
where $\tilde{\Psi}(\mathbf{r})$ is the real space Fourier transform of
$\Psi_{\mathbf{p}}$. Therefore, $\xi$ can be interpreted as the root-mean-square exciton
radius.

\subsection{Many-body scattering Hamiltonian\label{sub:hamiltonian}}

With exchange interactions ignored, the many-body scattering Hamiltonian for an exciton
immersed in an electron gas can be written as
\begin{eqnarray}
  \hat{\mathcal{H}}_{\text{eff}}
  &=&
  \sum_{\mathbf{K}}\varepsilon_{\text{X},\mathbf{K}}
  \hat{X}^{\dagger}_{\mathbf{K}}\hat{X}_{\mathbf{K}}
  +\sum_{\mathbf{k}}\varepsilon_{\text{e},\mathbf{k}}
  \hat{c}^{\dagger}_{\mathbf{k}}\hat{c}_{\mathbf{k}}\n
  &&+\sum_{\mathbf{Q}\mathbf{K}\mathbf{K}'}V_{\mathbf{K},\mathbf{K}'}
  \hat{c}^{\dagger}_{\mathbf{K}+\mathbf{Q}}\hat{c}_{\mathbf{K}'+\mathbf{Q}}
  \hat{X}^{\dagger}_{-\mathbf{K}}\hat{X}_{-\mathbf{K}'},
\end{eqnarray}
where the exciton kinetic energy and electron kinetic energy can be written as
\begin{eqnarray}
  \varepsilon_{\text{X},\mathbf{K}}
  =
  \varepsilon_{\text{X}}+\frac{|\mathbf{K}|^2}{2M_{\text{X}}},\hskip2ex
  \varepsilon_{\text{e},\mathbf{k}}
  =
  \frac{|\mathbf{k}|^2}{2m_{\text{e}}},
\end{eqnarray}
where $\varepsilon_{\text{X}}=\varepsilon_{\text{X},\mathbf{K}=\mathbf{0}}$ is the exciton
transition energy, $M_{\text{X}}$ is the exciton mass and $m_{\text{e}}$ is the electron
mass. The scattering potential is assumed to have the form
\begin{eqnarray}
  V_{\mathbf{K},\mathbf{K}'}
  &=&
  -\frac{v_{\mathbf{K}-\mathbf{K}'}}{L^2}
  \left[1-\exp\left(-\frac{1}{2}|\mathbf{K}-\mathbf{K}'|^2\xi^2\right)\right],
  \label{potential}
\end{eqnarray}
where $v_{\mathbf{K}-\mathbf{K}'}$ can be taken to be the screened Coulomb potential and
$\xi$ is the exciton radius.

The electric current operator for the electron-hole excitation process is given by
\begin{eqnarray}
  \hat{{j}}
  &\simeq&
  e\left(\mathcal{P}\hat{X}^{\dagger}_{\mathbf{0}}
  +\mathcal{P}^*\hat{X}_{\mathbf{0}}\right),
\end{eqnarray}
where $\mathcal{P}=\sum_{\mathbf{p}}\Psi_{\mathbf{p}}\mathcal{P}_{\mathbf{p}\text{cv}}$ is
the transition momentum matrix element. The absorption spectrum can be calculated by the
real part of the optical conductivity or the imaginary part of the current-current
response function\cite{mahan2013many},
\begin{eqnarray}
  \mathcal{A}(\omega)
  &=&
  2\;\text{Re }\sigma(\omega)
  =
  -\frac{2}{\omega}\text{Im }\int^{\infty}_{-\infty}
  \text{d}t\;e^{\mathtt{i}\omega t}\pi^{\text{R}}(t),\\
  \pi^{\text{R}}(t)
  &=&
  -\mathtt{i}\theta(t)\braket{\text{G}|\left[\hat{j}(t),\hat{j}\right]|\text{G}},
\end{eqnarray}
where $\theta(t)$ is a step function. The ground state
$\ket{\text{G}}=\ket{0}\ket{\text{FS}}$ is a direct product of the vacuum state of the
exciton ($\ket{0}$) and the Fermi sea of electrons ($\ket{\text{FS}}$). The response
function can be obtained from the exciton Green's function,
\begin{eqnarray}
  \pi^{\text{R}}(t)
  &=&
  e^2|\mathcal{P}|^2\mathcal{G}^{\text{R}}(t),
\end{eqnarray}
\begin{eqnarray}
  \mathcal{G}^{\text{R}}(t)
  &=&
  -\mathtt{i}\theta(t)\braket{\text{G}|\left[\hat{X}_{\mathbf{0}}(t),\;
  \hat{X}^{\dagger}_{\mathbf{0}}\right]|\text{G}}.
  \label{green_fun}
\end{eqnarray}
Then the absorption spectrum is expressed as
\begin{eqnarray}
  \mathcal{A}(\omega)
  &=&
  -\frac{2e^2}{\omega}|\mathcal{P}|^2\text{Im }\mathcal{G}^{\text{R}}(\omega).
\end{eqnarray}
The commutator in the Green's function written in Eq.~(\ref{green_fun}) implies that the
exciton operators are presumed to be bosonic particles. However, since the ground state is
an empty state for the exciton, and only single exciton generation is considered, the type
of the commutation relation chosen in Eq.~(\ref{green_fun}) does not alter the final
results.

\subsection{Scattering function and Fermi polaron theory\label{sub:scattering_function}}

In the frequency domain, the exciton Green's function can be solved by the Dyson's
equation
\begin{eqnarray}
  \mathcal{G}^{\text{R}}(\omega)
  &=&
  \frac{1}{\omega-\varepsilon_{\text{X}}-\Sigma^{\text{R}}(\omega)},
\end{eqnarray}
where $\Sigma^{\text{R}}(\omega)$ is the exciton self-energy. Based on perturbation
theory, the lowest-order expression for the exciton self-energy is given by
\begin{eqnarray}
  {\Sigma}^{\text{R}}(\omega)
  &=&
  \sum_{\mathbf{p}\mathbf{Q}}
  \frac{\left(1-n_{\mathbf{p}+\frac{m_{\text{e}}}{M_{\text{T}}}\mathbf{Q}}\right)
  n_{\mathbf{Q}}
  |V_{\mathbf{p}+\frac{m_{\text{e}}}{M_{\text{T}}}\mathbf{Q},\mathbf{Q}}|^2}
  {\omega-{\varepsilon}_{\text{X},-\mathbf{p}+\frac{M_{\text{X}}}{M_{\text{T}}}\mathbf{Q}}
  -{\varepsilon}_{\text{e},\mathbf{p}+\frac{m_{\text{e}}}{M_{\text{T}}}\mathbf{Q}}
  +\varepsilon_{\text{e},\mathbf{Q}}+\mathtt{i}0^+},\n
  \label{second_born}
\end{eqnarray}
where $M_{\text{T}}=M_{\text{X}}+m_{\text{e}}$ is the trion mass. It is not difficult to
show that the imaginary part of the self-energy corresponds to the damping constant of
electron-exciton scattering derived from Fermi's golden rule,
\begin{eqnarray}
  \eta_{\text{e-X}}
  &=&
  -2\text{Im }{\Sigma}^{\text{R}}(\varepsilon_{\text{X}}).
\end{eqnarray}
This confirms that the self-energy is simply the second-Born approximation of the
electron-exciton scattering problem. However, one cannot find a bound-state solution via a
simple perturbative self-energy approximation. One needs to consider, in an exact or
approximate way, the complete Born series. We assume that at lowest-order the two-particle
scattering function can be written as
\begin{eqnarray}
  \Gamma^{\text{R}}_{\mathbf{Q},\mathbf{Q}}(\omega)
  &=&
  \sum_{\mathbf{p}}\frac{\left(1-n_{\mathbf{p}
  +\frac{m_{\text{e}}}{M_{\text{T}}}\mathbf{Q}}\right)
  |V_{\mathbf{p}+\frac{m_{\text{e}}}{M_{\text{T}}}\mathbf{Q},\mathbf{Q}}|^2}
  {\omega-{\varepsilon}_{\text{X},-\mathbf{p}+\frac{M_{\text{X}}}{M_{\text{T}}}\mathbf{Q}}
  -{\varepsilon}_{\text{e},\mathbf{p}
  +\frac{m_{\text{e}}}{M_{\text{T}}}\mathbf{Q}}+\mathtt{i}0^+},\n
\end{eqnarray}
with the self-energy
\begin{eqnarray}
  \Sigma^{\text{R}}(\omega)
  &=&
  \sum_{\mathbf{Q}}n_{\mathbf{Q}}
  \Gamma^{\text{R}}_{\mathbf{Q},\mathbf{Q}}(\omega+\varepsilon_{\text{e},\mathbf{Q}}).
\end{eqnarray}
The scattering function can be extended to the solution of a two-particle
Lippmann-Schwinger equation\cite{lippmann1950variational, taylor2006scattering,
vagov2007generalized}
\begin{eqnarray}
  \Gamma^{\text{R}}_{\mathbf{Q},\mathbf{Q}'}(\omega)
  &=&
  V_{\mathbf{Q},\mathbf{Q}'}
  +\sum_{\mathbf{p}}V_{\mathbf{Q},\mathbf{p}+\frac{m_{\text{e}}}{M_{\text{T}}}\mathbf{Q}}\n
  &&\times\frac{\left(1-n_{\mathbf{p}+\frac{m_{\text{e}}}{M_{\text{T}}}\mathbf{Q}}\right)
  \Gamma_{\mathbf{p}+\frac{m_{\text{e}}}{M_{\text{T}}}\mathbf{Q},\mathbf{Q}'}(\omega)}
  {\omega-\varepsilon_{\text{X},-\mathbf{p}+\frac{M_{\text{X}}}{M_{\text{T}}}\mathbf{Q}}
  -\varepsilon_{\text{e},\mathbf{p}+\frac{m_{\text{e}}}{M_{\text{T}}}\mathbf{Q}}
  +\mathtt{i}0^+},\n
  \label{LSE1}
\end{eqnarray}
where we have used $V_{\mathbf{Q},\mathbf{Q}}=0$. By solving the Lippmann-Schwinger
equation, bound states can be found as the poles of the scattering function.

If we assume that for s-wave scattering the scattering potential and the scattering
function can be approximated as
\begin{eqnarray}
  V_{\mathbf{Q},\mathbf{Q}'}\simeq V,\hskip2ex
  \Gamma^{\text{R}}_{\mathbf{Q},\mathbf{Q}'}(\omega)
  \simeq\Gamma^{\text{R}}(\omega,\mathbf{Q}),
  \label{local_approx}
\end{eqnarray}
then the Lippmann-Schwinger equation becomes
\begin{eqnarray}
  \Gamma^{\text{R}}(\omega,\mathbf{Q})
  =
  V+VK^{\text{R}}(\omega,\mathbf{Q})\Gamma^{\text{R}}(\omega,\mathbf{Q}),
  \label{LSE2}
\end{eqnarray}
where the scattering kernel is
\begin{eqnarray}
  K^{\text{R}}(\omega,\mathbf{Q})
  &=&
  \sum_{\mathbf{p}}
  \frac{1-n_{\mathbf{p}+\frac{m_{\text{e}}}{M_{\text{T}}}\mathbf{Q}}}
  {\omega-\varepsilon_{\text{X},-\mathbf{p}+\frac{M_{\text{X}}}{M_{\text{T}}}\mathbf{Q}}
  -\varepsilon_{\text{e},\mathbf{p}+\frac{m_{\text{e}}}{M_{\text{T}}}\mathbf{Q}}
  +\mathtt{i}0^+}.\n
  \label{kernel}
\end{eqnarray}
The two energy terms that appear as poles of the kernel function can be rewritten as
\begin{eqnarray}
  {\varepsilon}_{\text{X},-\mathbf{p}+\frac{M_{\text{X}}}{M_{\text{T}}}\mathbf{Q}}
  +{\varepsilon}_{\text{e},\mathbf{p}+\frac{m_{\text{e}}}{M_{\text{T}}}\mathbf{Q}}
  &=&
  {\varepsilon}_{\text{X}}+\frac{|\mathbf{Q}|^2}{2M_{\text{T}}}
  +\frac{|\mathbf{p}|^2}{2\overline{m}_{\text{T}}},
  \label{trion_representation}
\end{eqnarray}
where $\overline{m}_{\text{T}}=\left(M^{-1}_{\text{X}}+m^{-1}_{\text{e}}\right)^{-1}$ is
the electron-exciton reduced mass. Eq.~(\ref{LSE2}) and Eq.~(\ref{kernel}) are effectively
the starting point of Efimkin and MacDonald's Fermi-polaron theory\cite{efimkin2017many}.
If the scattering process occurs in two dimensions, the approximation in
Eq.~(\ref{local_approx}) introduces a bound state with a binding energy depending on the
ultraviolate cut-off momentum, $k_{\Lambda}$, as\cite{efimkin2017many}
\begin{eqnarray}
  \Delta_{\text{T}}
  &=&
  \frac{k^2_{\Lambda}}{2\overline{m}_{\text{T}}}
  \exp\left({-\frac{2\pi}{\overline{m}_{\text{T}}g}}\right),
\end{eqnarray}
with the coupling constant $g$ connected to $V$ via $V=-g/L^2$. Note that the cut-off
energy can be related to the band-width of the conduction band and is thus not unphysical.
However, a cut-off dependent bound state is fundamentally inconsistent with the trion
model, valid in the limit of small doping, based on solving the three-particle
Schr\"{o}dinger equation. The cut-off dependent bound state originates from the contact
potential approximation in Eq.~(\ref{local_approx}), and the cut-off dependence is known
as the quantum anomaly in the two-dimensional quantum scattering
problem\cite{holstein1993anomalies, nyeo2000regularization}. To avoid this problem, an
interaction with some spatial range must be retained, such that a more sophisticated
method of solution is required.

\subsection{BSE and dynamical trion-hole generation\label{sub:BSE}}

To avoid the brute force solution of an integral equation, the two-particle
Lippmann-Schwinger equation can be solved by an orthogonal polynomial expansion, with
basis functions given by the eigenstates of the corresponding two-particle Schr\"{o}dinger
equation\cite{vagov2007generalized}. The connection between Lippmann-Schwinger equation
and Schr\"{o}dinger equation can be derived by a BSE formalism. The two-particle
Lippmann-Schwinger equation in Eq.~(\ref{LSE1}) can be rewritten as
\begin{eqnarray}
  \Gamma^{\text{R}}_{\mathbf{Q},\mathbf{Q}'}(\omega)
  &=&
  V_{\mathbf{Q},\mathbf{Q}'}+\sum_{\mathbf{p}\mathbf{p}'}
  V_{\mathbf{Q},\mathbf{p}+\frac{m_{\text{e}}}{M_{\text{T}}}\mathbf{Q}}\n
  &&\times
  P^{\text{R}}_{\mathbf{p},\mathbf{p}'}(\omega,\mathbf{Q})
  V_{\mathbf{p}'+\frac{m_{\text{e}}}{M_{\text{T}}}\mathbf{Q},\mathbf{Q}'},
  \label{LSE_BSE}
\end{eqnarray}
where the polarization is given by
\begin{eqnarray}
  P^{\text{R}}_{\mathbf{p},\mathbf{p}'}(\omega,\mathbf{Q})
  &=&
  \delta_{\mathbf{p},\mathbf{p}'}\Pi^{\text{R}}_{\mathbf{p}}(\omega,\mathbf{Q})\n
  &&+\sum_{\mathbf{q}}\Pi^{\text{R}}_{\mathbf{p}}(\omega,\mathbf{Q})
  V_{\mathbf{p},\mathbf{q}}P^{\text{R}}_{\mathbf{q},\mathbf{p}'}(\omega,\mathbf{Q}),
  \label{BSE}
\end{eqnarray}
and the zeroth-order polarization is
\begin{eqnarray}
  \Pi^{\text{R}}_{\mathbf{p}}(\omega,\mathbf{Q})
  &=&
  \frac{1-n_{\mathbf{p}+\frac{m_{\text{e}}}{M_{\text{T}}}\mathbf{Q}}}
  {\omega-\varepsilon_{\text{X},-\mathbf{p}+\frac{M_{\text{X}}}{M_{\text{T}}}\mathbf{Q}}
  -\varepsilon_{\text{e},\mathbf{p}+\frac{m_{\text{e}}}{M_{\text{T}}}\mathbf{Q}}
  +\mathtt{i}0^+}.
\end{eqnarray}
Eq.~(\ref{BSE}) is known as the BSE. The BSE can be solved by the spectral representation
\begin{eqnarray}
  P^{\text{R}}_{\mathbf{p},\mathbf{p}'}(\omega,\mathbf{Q})
  &=&
  \sum_{n}\frac{\rho_{n,\mathbf{Q}}
  \Phi_{\mathbf{p};n,\mathbf{Q}}\Phi^*_{\mathbf{p}';n,\mathbf{Q}}}
  {\omega-{\varepsilon}_{\text{X}}-\tilde{\varepsilon}_{n,\mathbf{Q}}+\mathtt{i}0^+},
  \label{polarization}
\end{eqnarray}
where $\rho_{n,\mathbf{Q}}$ is the distribution function of allowed transitions, and
$\tilde{\varepsilon}_{n,\mathbf{Q}}$ and $\Phi_{\mathbf{p};n,\mathbf{Q}}$ are obtained
from the two-particle Schr\"{o}dinger equation
\begin{eqnarray}
  &&\sum_{\mathbf{p}'}\Bigg[\delta_{\mathbf{p},\mathbf{p}'}
  \left(\frac{|\mathbf{Q}|^2}{2M_{\text{T}}}
  +\frac{|\mathbf{p}|^2}{2\overline{m}_{\text{T}}}\right)\n
  &&+V_{\mathbf{p}+\frac{m_{\text{e}}}{M_{\text{T}}}\mathbf{Q},
  \mathbf{p}'+\frac{m_{\text{e}}}{M_{\text{T}}}\mathbf{Q}}\Bigg]
  \Phi_{\mathbf{p}';n,\mathbf{Q}}
  =
  \tilde{\varepsilon}_{n,\mathbf{Q}}\Phi_{\mathbf{p};n,\mathbf{Q}}.
  \label{schrodinger_eq}
\end{eqnarray}
Note that Eq.~(\ref{trion_representation}) has been used to derive the kinetic-energy part
of the Schr\"{o}dinger equation. The distribution function of allowed transitions is
defined by
\begin{eqnarray}
  \rho_{n,\mathbf{Q}}
  &\equiv&
  \braket{\text{G}|\big\{\hat{T}_{n,\mathbf{Q}},\;
  \hat{T}^{\dagger}_{n,\mathbf{Q}}\big\}|\text{G}},
\end{eqnarray}
where $\hat{T}^{\dagger}_{n,\mathbf{Q}}=\sum_{\mathbf{p}}\Phi_{\mathbf{p};n,\mathbf{Q}}
\hat{X}^{\dagger}_{-\mathbf{p}+({M_{\text{X}}}/{M_{\text{T}}})\mathbf{Q}}
\hat{c}^{\dagger}_{\mathbf{p}+({m_{\text{e}}}/{M_{\text{T}}})\mathbf{Q}}$ is the trion
creation operator. Since the exciton state is initially empty, we find
$\braket{\hat{X}_{\mathbf{k}}\hat{X}^{\dagger}_{\mathbf{k}'}}
=\delta_{\mathbf{k},\mathbf{k}'}$ and
$\braket{\hat{X}^{\dagger}_{\mathbf{k}'}\hat{X}_{\mathbf{k}}}=0$. The distribution
function of allowed transitions becomes
\begin{eqnarray}
  \rho_{n,\mathbf{Q}}
  &=&
  \sum_{\mathbf{p}}
  \Phi^*_{\mathbf{p};n,\mathbf{Q}}\Phi_{\mathbf{p};n,\mathbf{Q}}
  \braket{\text{FS}|\hat{c}_{\mathbf{p}+\frac{m_{\text{e}}}{M_{\text{T}}}\mathbf{Q}}
  \hat{c}^{\dagger}_{\mathbf{p}+\frac{m_{\text{e}}}{M_{\text{T}}}\mathbf{Q}}|\text{FS}}\n
  &=&
  1-\sum_{\mathbf{p}}|\Phi_{\mathbf{p};n,\mathbf{Q}}|^2
  n_{\mathbf{p}+\frac{m_{\text{e}}}{M_{\text{T}}}\mathbf{Q}}.
\end{eqnarray}
This function allows for the bound state to be unfilled even though the transition energy
is below the Fermi energy, and introduces a Pauli-blocking effect to the transition as the
doping density increases. From Eq.~(\ref{LSE_BSE}) and Eq.~(\ref{polarization}), the
exciton self-energy can be solved as
\begin{eqnarray}
  \Sigma^{\text{R}}(\omega)
  &=&
  \sum_{\mathbf{Q}}\sum_{n}\frac{n_{\mathbf{Q}}\rho_{n,\mathbf{Q}}
  \mathcal{B}_{n,\mathbf{Q}}}
  {\omega-{\varepsilon}_{\text{X}}-\tilde{\varepsilon}_{n,\mathbf{Q}}
  +\varepsilon_{\text{e},\mathbf{Q}}+\mathtt{i}0^+},
  \label{BSE_self_energy}
\end{eqnarray}
where the self-energy spectral density is
\begin{eqnarray}
  \mathcal{B}_{n,\mathbf{Q}}
  &=&
  \Big\vert\sum_{\mathbf{p}}
  V_{\mathbf{Q},\mathbf{p}+\frac{m_{\text{e}}}{M_{\text{T}}}\mathbf{Q}}
  {\Phi}_{\mathbf{p};n,\mathbf{Q}}\Big\vert^2.
\end{eqnarray}
Therefore, the eigenvalues and eigenstates of Eq.~(\ref{schrodinger_eq}) contain all the
information needed to construct the self-energy. The eigenstates include the
electron-exciton bound states, which we interpret as the trion states, and the
electron-exciton scattering states.

The formalism described here is equivalent to Bronold's dynamical theory of trion-hole
generation\cite{bronold2000absorption}, except that Bronold solved for the trion
transition energy and the trion state by using a three-particle Schr\"{o}dinger equation.
Note that the exact solutions of Eq.~(\ref{schrodinger_eq}) is intrinsically independent
of the cut-off momentum. Since the BSE formalism is equivalent to the two-particle
Lippmann-Schwinger equation formalism, the theory of dynamical trion-hole generation is
consistent with the Fermi-polaron picture with a cut-off independent trion bound state.

\subsection{MND theory and Fermi-edge singularity\label{sub:MND}}

To introduce Fermi-edge singularity behavior, one possible method is to include the
self-energy corresponding to the contributions of multiple electron-hole pair excitations.
However, this method is complicated and numerically demanding. A simpler method is to
reduce the present formalism to an electron-exciton scattering version of the MND theory,
which was originally used to describe the response of the Fermi sea to a core-hole
potential\cite{mahan1967excitons, nozieres1969singularities, gavoret1969optical,
combescot1971infrared, mahan1980final, von1982dynamical}. The MND theory has a
straightforward and numerically exact solution for the response function in terms of a
time-dependent determinantal formulation. In this section, we will derive the MND
Hamiltonian and introduce this solution.

By a change of variables, the effective Hamiltonian can be reformulated as
\begin{eqnarray}
  \hat{\mathcal{H}}_{\text{eff}}
  &=&
  \sum_{\mathbf{Q}}\varepsilon_{\text{X},\mathbf{Q}}
  \hat{X}^{\dagger}_{\mathbf{Q}}\hat{X}_{\mathbf{Q}}
  +\sum_{\mathbf{k}}\varepsilon_{\text{e},\mathbf{k}}
  \hat{c}^{\dagger}_{\mathbf{k}}\hat{c}_{\mathbf{k}}\n
  &&+\sum_{\mathbf{Q}\mathbf{K}\mathbf{K}'}V_{\mathbf{K},\mathbf{K}'}
  \hat{c}^{\dagger}_{\mathbf{K}}\hat{c}_{\mathbf{K}'}
  \hat{X}^{\dagger}_{\mathbf{Q}-\mathbf{K}+\mathbf{K}'}\hat{X}_{\mathbf{Q}}.
\end{eqnarray}
Assuming that the exciton mass is infinitely large, the excitonic states can be expressed
as
\begin{eqnarray}
  \hat{X}^{\dagger}_{\mathbf{Q}-\mathbf{K}+\mathbf{K}'}\hat{X}_{\mathbf{Q}}
  &\simeq&
  \hat{X}^{\dagger}_{\mathbf{Q}}\hat{X}_{\mathbf{Q}}.
\end{eqnarray}
Due to the infinite exciton mass, the exciton transition energy becomes
$\varepsilon_{\text{X},\mathbf{Q}}\simeq\varepsilon_{\text{X}}$, and the quasi-momentum
$\mathbf{Q}$ becomes irrelevant. The effective Hamiltonian can be reduced to the MND
Hamiltonian\cite{mahan1967excitons, nozieres1969singularities}
\begin{eqnarray}
  \hat{\mathcal{H}}_{\text{MND}}
  &=&
  \varepsilon_{\text{X}}\hat{X}^{\dagger}\hat{X}
  +\sum_{\mathbf{k}}\varepsilon_{\text{e},\mathbf{k}}
  \hat{c}^{\dagger}_{\mathbf{k}}\hat{c}_{\mathbf{k}}
  +\sum_{\mathbf{k}\mathbf{k}'}V_{\mathbf{k},\mathbf{k}'}
  \hat{c}^{\dagger}_{\mathbf{k}}\hat{c}_{\mathbf{k}'}\hat{X}^{\dagger}\hat{X},\n
\end{eqnarray}
where $\hat{X}^{\dagger}$ and $\hat{X}$ are creation and annihilation operators of the
immobile exciton.

Based on this Hamiltonian, the exciton Green's function in the frequency-domain can be
solved exactly by\cite{baeten2015many, combescot1971infrared, mahan1980final,
von1982dynamical, schmidt2018universal}
\begin{eqnarray}
  \mathcal{G}^{\text{R}}(\omega)
  &=&
  -\mathtt{i}\int^{\infty}_{0}\text{d}t\;
  e^{\mathtt{i}\left(\omega-\varepsilon_{\text{X}}\right)t}
  \text{det}\big[{S}_{\mathbf{k},\mathbf{k}'}(t)
  \big]_{|\mathbf{k}|,|\mathbf{k}'|\leq k_{\text{F}}},\n
\end{eqnarray}
where $\text{det}[{S}_{\mathbf{k},\mathbf{k}'}(t)]_{|\mathbf{k}|,|\mathbf{k}'|\leq
k_{\text{F}}}$ is the determinant of the matrix
\begin{eqnarray}
  {S}_{\mathbf{k},\mathbf{k}'}(t)
  =
  \sum_{n}\Phi^*_{\mathbf{k},n}
  \exp\left[{-\mathtt{i}\left(\tilde{\varepsilon}_{n}
  -\varepsilon_{\mathbf{k}'}\right)t}\right]\Phi_{\mathbf{k}',n},
\end{eqnarray}
with $\tilde{\varepsilon}_{n}$, $\Phi_{\mathbf{k},n}$ given by the solution of the
eigenvalue problem
\begin{eqnarray}
  \sum_{\mathbf{k}'}\left(\delta_{\mathbf{k},\mathbf{k}'}
  \frac{|\mathbf{k}|^2}{2m_{\text{e}}}
  +V_{\mathbf{k},\mathbf{k}'}\right)\Phi_{\mathbf{k}',n}
  =
  \tilde{\varepsilon}_{n}\Phi_{\mathbf{k},n}.
  \label{MND_eigen}
\end{eqnarray}
The exciton Green's function can be generalized to include the effect of finite
temperatures and spectral line-broadening by including an electron distribution factor
($n_{\mathbf{k}}$) and a line-broadening parameter ($\gamma$) as\cite{baeten2015many,
schmidt2018universal}
\begin{eqnarray}
  \mathcal{G}^{\text{R}}(\omega)
  &=&
  -\mathtt{i}\int^{\infty}_{0}\text{d}t\;
  e^{\mathtt{i}\left(\omega-\varepsilon_{\text{X}}+\mathtt{i}\gamma\right)t}\n
  &&\times
  \text{det}\Big[\delta_{\mathbf{k},\mathbf{k}'}(1-n_{\mathbf{k}})+n_{\mathbf{k}}
  {S}_{\mathbf{k},\mathbf{k}'}(t)\Big].
\end{eqnarray}

\section{Exciton and trion transitions\label{sec:transitions}}

In this section, we scrutinize the exciton Green's function within the BSE formalism to
investigate the exciton and trion transitions in two dimensions. Since the exciton
self-energy contributes to the linewidth broadening of the exciton peak and the emerging
trion peak, an analytical study of the self-energy can help us understand the effects of
the electron-exciton interaction on the doping dependence of the optical spectra. In
Sec.~\ref{sub:trion}, the exciton self-energy and the bound-state solution of the BSE are
used to discuss the trion peak emergence. In Sec.~\ref{sub:exciton}, the scattering-state
solutions of the BSE is included to explain exciton linewidth broadening. In
Sec.~\ref{sub:oscillator}, the oscillator strength transfer from the exciton peak to the
trion peak is studied by analytically solving for the spectral weight of the exciton
transition. In Sec. V we follow up on our analytical explorations with extensive numerical
calculations.

\subsection{Trion peak emergence\label{sub:trion}}

Based on the sign of the eigenvalues of the two-particle Schr\"{o}dinger equation at
$\mathbf{Q}=\mathbf{0}$, the exciton self-energy can be separated into bound-state
contributions, $\Sigma^{\text{R}}_{\text{b}}(\omega)$ for
$\tilde{\varepsilon}_{n,\mathbf{0}}<0$, and scattering-state contributions,
$\Sigma^{\text{R}}_{\text{s}}(\omega)$, for $\tilde{\varepsilon}_{n,\mathbf{0}}\geq 0$. In
the following, we will illustrate that the bound-state contribution is responsible for the
emerging trion lineshape and that the electron-exciton scattering states contribute to the
exciton linewidth broadening and energy shift.

In the low-doping density regime, only $|{\mathbf{Q}}|\rightarrow{0}$ needs be considered.
The self-energy spectral density $B_{n,\mathbf{Q}}$ in this regime is insensitive to
$\mathbf{Q}$ and can be approximated as a constant. Assuming that the the lowest
eigenvalue of the Schr\"{o}dinger equation is
\begin{eqnarray}
  \tilde{\varepsilon}_{0,\mathbf{Q}}
  &=&
  -\Delta_{\text{T}}+\frac{|\mathbf{Q}|^2}{2M_{\text{T}}},
\end{eqnarray}
the self-energy contribution from bound states can be written as
\begin{eqnarray}
  \Sigma^{\text{R}}_{\text{b}}(\omega)
  &\simeq&
  \int\text{d}\varepsilon\;\frac{\mathcal{D}(\varepsilon)f(\varepsilon)
  \rho_{0}\mathcal{B}_{0}}
  {\omega-{\varepsilon}_{\text{X}}+\Delta_{\text{T}}+\left(1-m_{\text{e}}/M_{\text{T}}
  \right)\varepsilon+\mathtt{i}0^+},\n
\end{eqnarray}
where $\mathcal{D}(\varepsilon)$ is the density of states, $f(\varepsilon)$ is the
Fermi-Dirac distribution function, $\rho_{0}$ is the transition probability to the bound
state, $\mathcal{B}_{0}$ is the self-energy spectral density of the bound state with
$|{\mathbf{Q}}|\rightarrow{0}$, and $\Delta_{\text{T}}$ is the trion binding energy. For a
two-dimensional system, $\mathcal{D}(\varepsilon)=\theta(\varepsilon)m_{\text{e}}/(2\pi)$.
In the zero-temperature limit,
$f(\varepsilon)=\theta(\varepsilon_{\text{F}}-\varepsilon)$, and the self-energy becomes
\begin{eqnarray}
  \Sigma^{\text{R}}_{\text{b}}(\omega)
  &\simeq&
  \alpha_0
  \ln\left[1+\frac{\left(1-m_{\text{e}}/M_{\text{T}}\right)\varepsilon_{\text{F}}}
  {\omega-\varepsilon_{\text{X}}+\Delta_{\text{T}}}\right]\n
  &&-\mathtt{i}\pi\alpha_0
  \theta\left({\varepsilon}_{\text{X}}-\Delta_{\text{T}}-\omega\right)\n
  &&\times\theta\left[\omega-{\varepsilon}_{\text{X}}+\Delta_{\text{T}}
  +\left(1-\frac{m_{\text{e}}}{M_{\text{T}}}\right)\varepsilon_{\text{F}}\right],
  \label{approx_SE}
\end{eqnarray}
where $\alpha_0={\rho_{0}\mathcal{B}_{0}m_{\text{e}}}/\left[{2\pi}
\left(1-{m_{\text{e}}}/{M_{\text{T}}}\right)\right]^{-1}$. The self-energy diverges at
$\omega=\varepsilon_{\text{X}}-\Delta_{\text{T}}$ and
$\omega=\varepsilon_{\text{X}}-\Delta_{\text{T}}
-\left(1-m_{\text{e}}/M_{\text{T}}\right)\varepsilon_{\text{F}}$ with
$\varepsilon_{\text{F}}>0$. The trion transition energy ($\varepsilon_{\text{T}}$) can be
solved from $\varepsilon_{\text{T}}-\varepsilon_{\text{X}}-\text{Re
}\Sigma^{\text{R}}(\varepsilon_{\text{T}})=0$ and turns out to be slightly larger than
$\varepsilon_{\text{X}}-\Delta_{\text{T}}
-\left(1-m_{\text{e}}/M_{\text{T}}\right)\varepsilon_{\text{F}}$. The linewidth broadening
arises from the imaginary part of the self-energy, which includes the branch-cut of the
first term and the whole second term in Eq.~(\ref{approx_SE}). The linewidth of the trion
peak is approximately
\begin{eqnarray}
  \eta_{\text{T}}
  =
  -2\text{Im }\Sigma^{\text{R}}_{\text{b}}(\varepsilon_{\text{T}})
  =
  2\pi\alpha_{0}.
\end{eqnarray}
In this regime, the trion transition is a discrete quasi-particle-like excitation as
opposed to a collective excitation, and the trion linewidth is basically independent of
the doping density.

\subsection{Exciton linewidth broadening\label{sub:exciton}}

In addition to the bound-state solutions of the BSE, the scattering-state solutions are
also included in the exciton self-energy. The transition energies of the scattering states
are close to or larger than the exciton transition energy, and the associated
wavefunctions can be approximated as plane-wave functions. For these scattering-state
contributions, the second-Born self-energy in Eq.~(\ref{second_born}) can be used as an
approximation. By a change of variables, Eq.~(\ref{second_born}) can be rewritten as
\begin{eqnarray}
  \Sigma^{\text{R}}_{\text{s}}(\omega)
  &=&
  \sum_{\mathbf{p}\mathbf{q}}\frac{\left(1-n_{\mathbf{p}+\mathbf{q}}\right)n_{\mathbf{q}}
  |V_{\mathbf{p}+\mathbf{q},\mathbf{q}}|^2}{\omega-\varepsilon_{\text{X}}
  -\frac{|\mathbf{p}|^2}{2M_{\text{X}}}
  -\frac{|\mathbf{p}+\mathbf{q}|^2}{2m_{\text{e}}}+\frac{|\mathbf{q}|^2}{2m_{\text{e}}}
  +\mathtt{i}0^+}.\n
\end{eqnarray}
Replacing the scattering potential by
$V_{\mathbf{p}+\mathbf{q},\mathbf{q}}=\tilde{v}_{\mathbf{p}}/L^2$ and converting summation
to integration, the self-energy can be reformulated as
\begin{eqnarray}
  \Sigma^{\text{R}}_{\text{s}}(\omega)
  &=&
  \int\frac{\text{d}^2\mathbf{p}\text{d}^2\mathbf{q}}{(2\pi)^4}
  \frac{\left(1-n_{\mathbf{p}+\mathbf{q}}\right)n_{\mathbf{q}}\tilde{v}^2_{\mathbf{p}}}
  {\omega-\varepsilon_{\text{X}}-\frac{|\mathbf{p}|^2}{2\overline{m}_{\text{T}}}
  -\frac{\mathbf{p}\cdot\mathbf{q}}{m_{\text{e}}}+\mathtt{i}0^+}.
\end{eqnarray}
In the integration, the range of quasi-momentum $\mathbf{q}$ is confined by the
electron-density distribution $n_{\mathbf{q}}$, and the range of quasi-momentum
$\mathbf{p}$ is determined by the electron-exciton potential $\tilde{v}_{\mathbf{p}}$.
Clearly, the exciton linewidth broadening is dependent on the form of the electron-exciton
potential.

For example, consider the case of a contact potential,
$\tilde{v}_{\mathbf{p}}=\tilde{v}_{0}$. Since the Fermi energy is generally smaller than
the bandwidth of conduction band, the range of $|\mathbf{q}|$ is much smaller than the
range of $|\mathbf{p}|$, such that we can ignore the term
$\mathbf{p}\cdot\mathbf{q}/m_{\text{e}}$ in the denominator and approximate
$n_{\mathbf{p}+\mathbf{q}}\simeq n_{\mathbf{p}}$. The self-energy becomes
\begin{eqnarray}
  \Sigma^{\text{R}}_{\text{s}}(\omega)
  &\simeq&
  n_{\text{D}}\tilde{v}^2_{0}\int\frac{\text{d}^2\mathbf{p}}{(2\pi)^2}
  \frac{\left(1-n_{\mathbf{p}}\right)}
  {\omega-\varepsilon_{\text{X}}-\frac{|\mathbf{p}|^2}{2\overline{m}_{\text{T}}}
  +\mathtt{i}0^+},
  \label{low_density_approx}
\end{eqnarray}
where $n_{\text{D}}=\int\text{d}^2\mathbf{q}n_{\mathbf{q}}/(2\pi)^2$ is the doping
density. The exciton linewidth broadening is then given by
\begin{eqnarray}
  \eta_{\text{X}}(\omega)
  &=&
  -2\text{Im }{\Sigma}^{\text{R}}_{\text{s}}(\omega)\n
  &\simeq&
  2\pi\;n_{\text{D}}\tilde{v}^2_{0}
  \int^{k_{\Lambda}}_{k_{\text{F}}}\frac{p\text{d}{p}}{2\pi}
  \delta\left(\omega-\varepsilon_{\text{X}}
  -\frac{p^2}{2\overline{m}_{\text{T}}}\right)\n
  &\propto&
  n_{\text{D}}\tilde{v}^2_{0}\;\theta\left(\omega-\varepsilon_{\text{X}}
  -\frac{m_{\text{e}}}{\overline{m}_{\text{T}}}\varepsilon_{\text{F}}\right).
\end{eqnarray}
Based on this approximate linewidth function, the exciton linewidth is proportional to the
doping density, and the lineshape is an asymmetric peak with a threshold energy
$\omega_{\text{TH}}=\varepsilon_{\text{X}}
+(m_{\text{e}}/\overline{m}_{\text{T}})\varepsilon_{\text{F}}$. On the other hand, if the
electron-exciton potential diverges as $|\mathbf{p}|\rightarrow{0}$, the approximation in
Eq.~(\ref{low_density_approx}) is no longer valid. As an extreme example, consider a case
where the scattering potential diverges at a quasi-momentum much smaller than
$k_{\text{F}}$; here we can assume $\tilde{v}_{\mathbf{p}}\sim
2\pi\tilde{v}_{0}\delta_{\mathbf{p},\mathbf{0}}$. The exciton linewidth broadening becomes
\begin{eqnarray}
  \eta_{\text{X}}(\omega)
  &=&
  -2\tilde{v}^2_{0}\;\text{Im}\int\frac{\text{d}^2\mathbf{q}}{(2\pi)^2}
  \frac{\left(1-n_{\mathbf{q}}\right)n_{\mathbf{q}}}
  {\omega-\varepsilon_{\text{X}}+\mathtt{i}0^+}\n
  &\sim&
  2\pi\tilde{v}^2_{0}\beta^{-1}\delta(\omega-\varepsilon_{\text{X}}).
\end{eqnarray}
In this case exciton linewidth broadening becomes independent of doping density and
proportional to the temperature. The exciton peak is also fixed at the vertical transition
energy.

Based on the above considerations, the doping-dependent exciton linewidth in Efimkin and
MacDonald's work\cite{efimkin2017many} and the doping-independent exciton linewidth in
Baeten and Wouters' work can be explained, since the former's work uses a contact
potential and the later's model uses a Yukawa potential\cite{baeten2015many} which
approaches singular behavior near zero quasi-momentum. Clearly the form of the
doping dependence is affected by the screening length of the Coulomb potential. A
numerical calculation is needed in realistic cases, as we will study in
Sec.~\ref{sub:doping}.

\subsection{Oscillator strength transfer\label{sub:oscillator}}

Based on standard Green's function considerations, the spectral weight of the exciton
transition can be calculated from\cite{mahan2013many}
\begin{eqnarray}
  Z_{\text{X}}
  &=&
  \left[1-\frac{\partial}{\partial\omega}\text{Re }\Sigma^{\text{R}}(\omega)
  \Big\vert_{\omega=\varepsilon_{\text{X}}+\Sigma^{\text{R}}(\varepsilon_{\text{X}})}
  \right]^{-1},
\end{eqnarray}
where $\Sigma^{\text{R}}(\omega)=\Sigma^{\text{R}}_{\text{b}}(\omega)
+\Sigma^{\text{R}}_{\text{s}}(\omega)$. Since the scattering-state self-energy only
contributes to the exciton line-broadening and does not affect the total area of the
lineshape, we only need consider the bound-state self-energy at the exciton transition
energy, $\varepsilon_{\text{X}}$, to probe the spectral weight shift. Thus the spectral
weight can be approximated as
\begin{eqnarray}
  Z_{\text{X}}
  &\simeq&
  \left[1-\frac{\partial}{\partial\omega}\text{Re }\Sigma^{\text{R}}_{\text{b}}(\omega)
  \Big\vert_{\omega=\varepsilon_{\text{X}}}\right]^{-1}\n
  &=&
  \Bigg[1+\frac{\alpha_{0}}{\Delta_{\text{T}}}
  \frac{\left(1-m_{\text{e}}/M_{\text{T}}\right)\varepsilon_{\text{F}}}
  {\Delta_{\text{T}}+\left(1-m_{\text{e}}/M_{\text{T}}\right)
  \varepsilon_{\text{F}}}\Bigg]^{-1}.
\end{eqnarray}
We find $Z_{\text{X}}=1$ at $\varepsilon_{\text{F}}=0$. Assuming that
$Z_{\text{X}}\rightarrow{0}$ as $\varepsilon_{\text{F}}\rightarrow\infty$, which implies
$\alpha_{0}/\Delta_{\text{T}}\rightarrow\infty$, the competition between exciton and trion
oscillator strengths would be significant at the scale of the Fermi energy
$\varepsilon_{\text{F}}\sim\Delta_{\text{T}}$. This matches the observed scale where the
exciton peak is depleted.

We note that $\alpha_0$ is a constant only for low doping-density regime, and thus all
statements of this section are restricted by this consideration. A full numerical
investigation will be carried out in Sec.~\ref{sub:doping}.

\section{Numerical calculations\label{sec:numerical}}

In this section, numerical calculations employing both the BSE formalism and the MND
theory are performed to compare the two methods and discuss the effects of a broad range
of doping densities as well as edge-singularity effects. First, it is necessary to discuss
the optical energy renormalization due to doping, since it will affect the positions of
the trion and exciton peaks. In Sec.~\ref{sub:2d_materials}, the parameters for
two-dimensional materials are given, and some energy-scale and momentum-scale parameters
of the two-dimensional electron gas model are introduced for usage in the following
discussions. In Sec.~\ref{sub:renormalization}, the scale of optical energy
renormalization due to doping electrons is estimated. In Sec.~\ref{sub:DOS}, the
differential density of states is plotted by solving the Schr\"{o}dinger equation with the
scattering potential. The calculated bound-state behavior and the relationship between the
trion peak calculated via the BSE formalism and the MND theory are discussed. In
Sec.~\ref{sub:doping} and Sec.~\ref{sub:temperature}, the doping dependence and
temperature dependence of the optical spectra calculated by BSE formalism and MND theory
are presented and discussed.

\subsection{2D materials\label{sub:2d_materials}}

The exciton transition energy of a doped material can be defined by the optical gap in the
absence of doping and the doping-induced energy shift,
\begin{eqnarray}
  \varepsilon_{\text{X}}=\Delta+\delta\Delta,
\end{eqnarray}
where $\Delta$ is the optical gap and $\delta\Delta$ is the optical energy renormalization
due to electron doping. For the BSE calculations, we assume that the electron mass and
hole mass are equivalent, $m_{\text{e}}=m_{\text{h}}$, such that the exciton mass is
$M_{\text{X}}=2m_{\text{e}}$, the trion mass is $M_{\text{T}}=3m_{\text{e}}$, and the
electron-exciton reduced mass is $\overline{m}_{\text{T}}=2m_{\text{e}}/3$. On the other
hand, for the MND calculations, the hole mass, exciton mass and trion mass are infinitely
large. The electron mass and electron-exciton reduced mass are equal
($m_{\text{e}}=\overline{m}_{\text{T}}$). We assume that the screened Coulomb potential in
two dimensions is described by the Rytova-Keldysh potential\cite{berkelbach2013theory,
rytova, Keldysh}. Via Eq.~(\ref{potential}), the scattering potential is written as
\begin{eqnarray}
  V_{\mathbf{k},\mathbf{k}'}
  &=&
  \frac{-2\pi e^2/L^2}{|\mathbf{k}-\mathbf{k}'|\left(1+r_0|\mathbf{k}-\mathbf{k}'|\right)}
  \left(1-e^{-|\mathbf{k}-\mathbf{k}'|^2\xi^2/2}\right),\n
\end{eqnarray}
where $L^2$ is the dimensional area, $r_0$ is the screening length, and $\xi$ is the
exciton radius. We use a finite-size square box with square-lattice points to approach the
infinite two-dimensional limit. The lattice k-grid is given by
\begin{eqnarray}
  \mathbf{k}=(k_x,k_y)
  =
  \left(\frac{2\pi}{L}\kappa_x,\frac{2\pi}{L}\kappa_y\right),
\end{eqnarray}
where $\kappa_x,\kappa_y=0,1,\cdots,N-1$ with $N$ the number of grid point in one
direction. The box dimensional length is $L=Na_{\Lambda}$, where $a_{\Lambda}$ is the
cut-off length and also the lattice constant. The cut-off momentum is
$k_{\Lambda}=2\pi/a_{\Lambda}$ and cut-off energy is
$\varepsilon_{\Lambda}=k_{\Lambda}^2/(2m_{\text{e}})$. In our calculations, we use the
parameters $\Delta=2.0$ eV, $\varepsilon_{\Lambda}=2.0$ eV, $r_0=36$ \AA, and the
effective electron mass is assumed to be
$m_{\text{e}}=0.045\;\text{eV}^{-1}\text{\AA}^{-2}$. We will discuss the dependence of
electron mass on the trion binding energy in Sec~\ref{sub:DOS}. The exciton radius $\xi$
is an adjustable parameter.

Electron doping in two-dimensional materials can be modeled by a noninteracting
two-dimensional electron gas. The electron distribution is given by
\begin{eqnarray}
  n_{\mathbf{k}}
  &=&
  \left[e^{\beta\left(\varepsilon_{\text{e},\mathbf{k}}-\varepsilon_{\text{F}}\right)}
  +1\right]^{-1},
\end{eqnarray}
where $\beta$ is the inverse temperature and $\varepsilon_{\text{F}}$ is the Fermi energy.
The total doping density is given by
\begin{eqnarray}
  n_{\text{D}}
  &=&
  \frac{\nu}{L^2}\sum_{\mathbf{k}}n_{\mathbf{k}}
  =
  \nu\int^{\infty}_{0}\frac{\text{d}k^2}{4\pi}
  \left[e^{\beta\left(k^2/(2m_{\text{e}})-\varepsilon_{\text{F}}\right)}
  +1\right]^{-1}\n
  &=&
  \int^{\infty}_{-\infty}\text{d}\varepsilon\;\mathcal{D}(\varepsilon)f(\varepsilon),
\end{eqnarray}
where $\nu$ is the degeneracy factor, $\mathcal{D}(\varepsilon)=\theta(\varepsilon){\nu
m_{\text{e}}}/({2\pi})$ is the density of states and
$f(\varepsilon)=1/\left[e^{\beta\left(\varepsilon-\varepsilon_{\text{F}}\right)}
+1\right]$ is the Fermi-Dirac distribution function. We can define the Fermi wavevector by
\begin{eqnarray}
  n_{\text{D}}
  &=&
  \nu\int^{k_{\text{F}}}_{0}\frac{k\text{d}k}{2\pi}
  =
  \frac{\nu k^2_{\text{F}}}{4\pi}.
\end{eqnarray}
The Fermi wavevector is given by $k_{\text{F}}=\sqrt{4\pi n_{\text{D}}/\nu}$. The chemical potential is defined as
\begin{eqnarray}
  \mu=\frac{k^2_{\text{F}}}{2m_{\text{e}}}.
\end{eqnarray}
Note that the chemical potential can be different from the Fermi energy at non-zero
temperatures and becomes equivalent to it, namely $\mu=\varepsilon_{\text{F}}$, at zero
temperature.

\subsection{Optical energy renormalization\label{sub:renormalization}}

The contributions to the optical energy renormalization include a Pauli-blocking effect
($\delta\Delta_{\text{PB}}$), a vertical-excitation shift ($\delta\Delta_{\text{VE}}$),
and a band-gap renormalization ($\delta\Delta_{\text{BG}}$)\cite{liu2016engineering,
yao2017optically}
\begin{eqnarray}
  \delta\Delta
  =
  \delta\Delta_{\text{PB}}+\delta\Delta_{\text{VE}}+\delta\Delta_{\text{BG}}.
\end{eqnarray}
There are two additional contributions which are often mentioned in the literature but we
do not consider here. One is the exciton binding energy renormalization and the other
arises from dynamical screening. The former increases the optical energy and the later
reduces it. According to some reports, the two terms are minor effects and roughly cancel
with each other\cite{zimmermann1978dynamical, schmitt1985excitons}. It should however be
noted that these studies are performed in the low doping density regime. In the remainder
of this work, we assume that these two contributions can be ignored.

The Pauli-blocking effect is assumed to be given by
\begin{eqnarray}
  \delta\Delta_{\text{PB}}=\mu,
\end{eqnarray}
since the conduction band states lower than the chemical potential are filled. The
vertical excitation shift is accompanied by a Pauli-blocking effect, since the the
quasi-momentum of the excited electron must be larger than the Fermi momentum
$k_{\text{F}}$ when the excited hole has the same quasi-momentum as the excited electron.
The energy shift is
\begin{eqnarray}
  \delta\Delta_{\text{VE}}
  &=&
  \frac{k^2_{\text{F}}}{2m_{\text{h}}}=\frac{m_{\text{e}}}{m_{\text{h}}}\mu.
\end{eqnarray}
The band-gap renormalization due to electron-electron interactions is
\begin{eqnarray}
  \delta{\Delta}_{\text{BG}}
  =
  \text{Re }\sigma_{\mathbf{k}}(\omega)\big\vert_{k=0,\;\omega=0},
\end{eqnarray}
where $\sigma_{\mathbf{k}}(\omega)$ is the quasi-particle self-energy. The self-energy is
given by the static screened-exchange approximation\cite{inkson1976effect,
gao2017renormalization} with a band-gap renormalization that can be written as
\begin{eqnarray}
  \delta{\Delta}_{\text{BG}}
  &=&
  -\frac{1}{L^2}\sum_{\mathbf{q}}n_{\mathbf{q}}W_{\mathbf{q}}.
\end{eqnarray}
The screened potential is given by
$W_{\mathbf{q}}=\left(v^{-1}_{\mathbf{q}}-\Pi_{\mathbf{q}}\right)^{-1}$, where the
Rytova-Keldysh potential $v_{\mathbf{q}}$ can be approximated as Coulomb potential in
long-wavelength regime ($q\leq{k}_{\text{F}}$)
\begin{eqnarray}
  {v}_{\mathbf{q}}\simeq\frac{2\pi e^2}{q}.
\end{eqnarray}
The polarization is given by the Stern's formula\cite{stern1967polarizability,
galitski2004universal, zhang2005quasiparticle}
\begin{eqnarray}
  \Pi_{\mathbf{q}}
  &=&
  -\frac{\nu m_{\text{e}}}{2\pi}\left[1-\theta\left(q-2k_{\text{F}}\right)
  \sqrt{1-\left(2k_{\text{F}}/q\right)^2}\right].
\end{eqnarray}
Since $\Pi_{\mathbf{q}}=-{\nu m_{\text{e}}}/({2\pi})$ when $q\leq{k}_{\text{F}}$, the
screened potential can be approximated as
\begin{eqnarray}
  W_{\mathbf{q}}\simeq\frac{2\pi e^2}{q+q_{\text{TF}}},
\end{eqnarray}
where $q_{\text{TF}}=\nu m_{\text{e}}e^2$ is the Thomas-Fermi wavevector in two
dimensions. The band-gap renormalization in the zero-temperature limit is given by
\begin{eqnarray}
  \delta{\Delta}_{\text{BG}}
  &=&
  -\int\frac{\text{d}^2\mathbf{q}}{(2\pi)^2}n_{\mathbf{q}}W_{\mathbf{q}}\n
  &\simeq&
  -e^2\left[k_{\text{F}}-q_{\text{TF}}
  \ln\left(1+\frac{k_{\text{F}}}{q_{\text{TF}}}\right)\right].
\end{eqnarray}
Since $k_{\text{F}}\ll q_{\text{TF}}$, we find
\begin{eqnarray}
  \delta{\Delta}_{\text{BG}}
  &\simeq&
  -e^2\left[k_{\text{F}}-q_{\text{TF}}
  \left(\frac{k_{\text{F}}}{q_{\text{TF}}}
  -\frac{k^2_{\text{F}}}{2q^2_{\text{TF}}}+\cdots\right)\right]\n
  &\simeq&
  -\frac{e^2k^2_{\text{F}}}{2q_{\text{TF}}}=-\frac{k^2_{\text{F}}}{2\nu m_{\text{e}}}
  =-\frac{\mu}{\nu}.
\end{eqnarray}
For a spin-degenerate electron gas, the band-gap renormalization is about
$\delta\Delta_{\text{BG}}\simeq-\mu/2$ as found from the static screened-exchange
approximation. However, for the case where the spin-orbit coupling is large enough to
split the spin degeneracy and the splitting energy is larger than or close to the chemical
potential, the degeneracy factor becomes $\nu=1$. In this case the value of the band-gap
renormalization becomes $\delta\Delta_{\text{BG}}\simeq-\mu$. This is the situation that
occurs in TMDC monolayers and in most semiconducting quantum wells. In the present
calculations, we will consider this later case.

In summary, when the contributions from band-gap renormalization and Pauli-blocking cancel
with each other, the total optical energy renormalization is approximately
\begin{eqnarray}
  \delta\Delta=\frac{m_{\text{e}}}{m_{\text{h}}}\mu,
\end{eqnarray}
and the optical energy renormalization for the BSE calculations carried out here is given
by $\delta\Delta=\mu$, since we choose equal electron and hole masses. The optical energy
renormalization for the present MND calculations is $\delta\Delta=0$, since the hole mass
is infinitely large.

\subsection{Electron-exciton scattering and bound states\label{sub:DOS}}

\begin{figure}
  \includegraphics[width=0.95\linewidth]{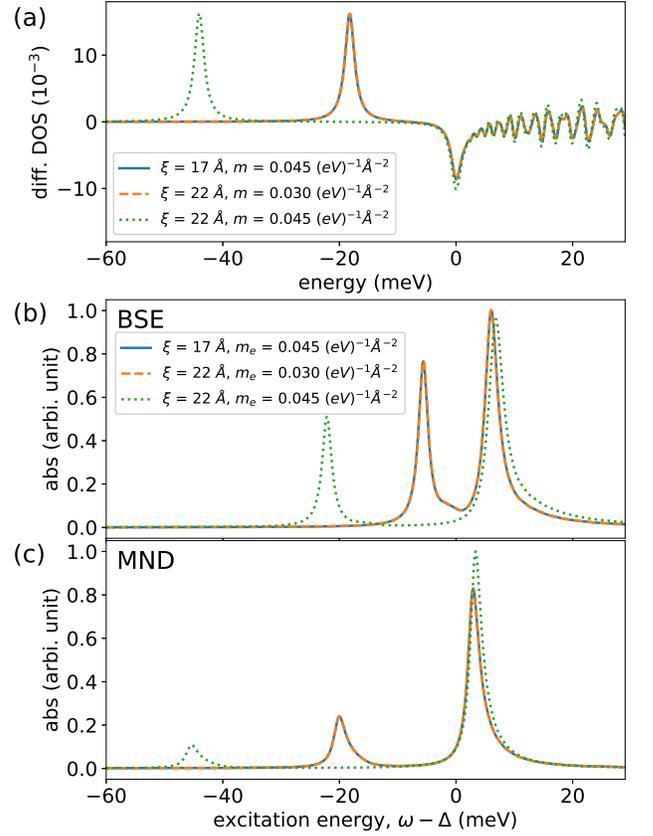}
  \caption{(a) Differential DOS; (b) Absorption spectrum from the BSE formalism; (c)
  Absorption spectrum from the MND theory with $\beta^{-1}=0$ meV,
  $\varepsilon_{\text{F}}=5$ meV, $\gamma=1$ meV, $N=140$.}
  \label{fig:DOS_spectra}
\end{figure}

In this section, we study the bound-state solution of the electron-exciton scattering
problem by solving the eigenvalue problem
\begin{eqnarray}
  \sum_{\mathbf{k}'}\left(\delta_{\mathbf{k},\mathbf{k}'}
  \frac{|\mathbf{k}|^2}{2m}+V_{\mathbf{k},\mathbf{k}'}\right)\Phi_{\mathbf{k}',n}
  =
  \tilde{\varepsilon}_{n}\Phi_{\mathbf{k},n}.
  \label{eigen_eq}
\end{eqnarray}
The equation is identical to Eq.~(\ref{MND_eigen}) in our discussion of the MND theory for
$m=m_{\text{e}}$. The equation with $m=\overline{m}_{\text{T}}$ can also be used to find
the eigenvalue and the eigenvector of Eq.~(\ref{schrodinger_eq}) since the center-of-mass
momentum can be decoupled as
\begin{eqnarray}
  \tilde{\varepsilon}_{n,\mathbf{Q}}
  &=&
  \tilde{\varepsilon}_{n}+\frac{|\mathbf{Q}|^2}{2M_{\text{T}}},\hskip2ex
  \Phi_{\mathbf{k};n,\mathbf{Q}}=\Phi_{\mathbf{k},n}.
\end{eqnarray}
In order to describe the impurity-induced bound states of the system, we define the
differential density of states (diff.\,DOS) as
\begin{eqnarray}
  \Delta\rho(\omega)
  &\equiv&
  \frac{1}{N^2}\left[
  \sum_{n}\delta(\omega-\tilde{\varepsilon}_{n})
  -\sum_{\mathbf{k}}\delta(\omega-\varepsilon_{\mathbf{k}})\right].
\end{eqnarray}
This quantity measures the spectral density shift from the non-interacting Fermi gas to
the reorganized Fermi gas due to the electron-exciton interaction. In
Fig.~\ref{fig:DOS_spectra} (a), the calculated diff.\,DOS with different values of the
electron mass and exciton radius is shown. It is found that a negative energy weakly bound
state exists for each parameter set. In the language employed in this work, the
weakly bound state is assigned as the trion state. The trion binding energy depends on the
mass and the exciton radius. It is found that the trion binding energy increases as the
electron mass and exciton radius parameters increase.

In Fig.~\ref{fig:DOS_spectra} (b), (c), the optical spectra given by (b) the BSE formalism
and (c) the MND theory with electron mass $m_{\text{e}}=m$ are shown for different exciton
radius. The peaks close to $\omega-\Delta\simeq{0}$ are assigned as exciton transitions
and the additional peaks with lower energies are assigned as trion transitions. As can be
seen, the trion binding energies calculated with the same parameters but by different
methods are quite different. The binding energy calculated by the BSE formalism is
consistently smaller than the one calculated by the MND theory. This occurs because that
the binding energy given by the BSE formalism is calculated by diagonalizing
Eq.~(\ref{eigen_eq}) with $m=\overline{m}_{\text{T}}=2m_{\text{e}}/3$, while the binding
energy given by the MND theory is calculated from the same equation with $m=m_{\text{e}}$.
The binding energy obtained from the BSE formalism with
$m_{\text{e}}=0.045\;(\text{eV})^{-1}\text{\AA}^{-2}$ is close to the binding energy of
the MND theory with $m_{\text{e}}=0.030\;(\text{eV})^{-1}\text{\AA}^{-2}$. Since the
electron mass can be found experimentally and is about
$m_{\text{e}}=0.045\;(\text{eV})^{-1}\text{\AA}^{-2}$, we choose different exciton radii
for different methods ($\xi=22$ \AA\; for BSE and $\xi=17$ \AA\; for MND) to adjust the
binding energies to similar values. Through this adjustment, we can compare the optical
spectra by the two methods and exclude the binding energy difference, thus enabling
consideration of edge-singularity effects.

Note that we use an overestimate of the exciton radius for each method and still find an
underestimate of the trion binding energy. For a monolayer TMDC, such as $\text{MoS}_2$,
the exciton radius is about $10$ \AA\; and the trion binding energy is over $20$ meV based
on the calculation of the three-particle Schr\"{o}dinger equation
model\cite{berkelbach2013theory}. The deviation is caused by some factors not considered
in the present theory. First, we do not include the exchange energy in our model. The
exchange energy can be described as the indistinguishability between the doping electron
and the bound electron in the exciton. For the Schr\"{o}dinger equation model, the
exchange energy can be included by assuming that the variational wavefunction is symmetric
to the exchange of the two electron degrees of freedom. Secondly, we do not consider the
effect whereby the exciton transition energy relaxes to lower values due to the
interaction with the doping electrons. Lastly, the electron-exciton scattering potential
has been written by assuming that the exciton wavefunction is a 1s-orbital solution of the
two-dimensional hydrogen atom. However, it is possible that there exists excited state
orbitals that hybridize with the exciton wavefunction when the exciton interacts with
doping electrons. All of these factors contribute to the observed deviations and are
beyond the present discussion. We will leave these issues to a future study.

\subsection{Doping-dependent optical spectra\label{sub:doping}}

\begin{figure}
  \includegraphics[width=0.95\linewidth]
  {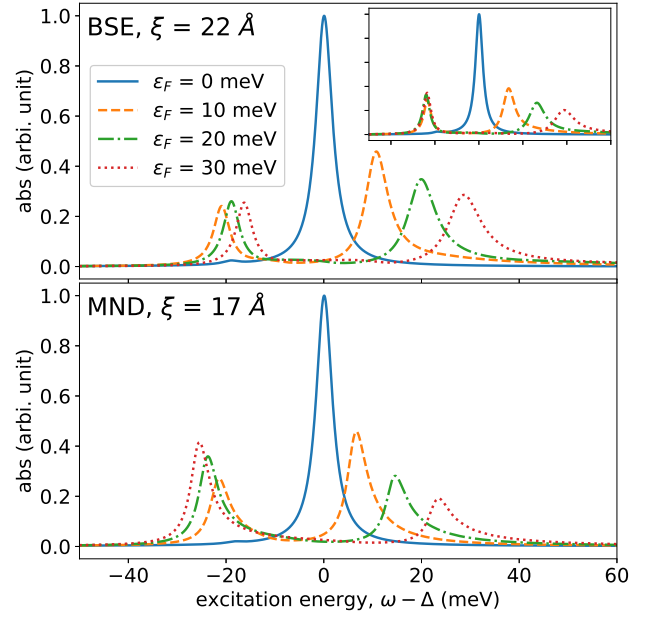}
  \caption{The doping-dependent optical spectra calculated by the BSE formalism (upper
  panel) and the MND theory (lower panel) with
  $m_{\text{e}}=0.045\;(\text{eV})^{-1}\text{\AA}^{-2}$, $\beta^{-1}=0.1$ meV, $\gamma=2$
  meV, $N=140$. The inset in upper panel shows the spectra calculated by the BSE formalism
  without the distribution function of allowed transition $\rho_{n,\mathbf{Q}}$ for $n=0$
  (the trion bound state) in Eq.~(\ref{BSE_self_energy}).}
  \label{fig:doping_dependent_spectra}
\end{figure}

\begin{figure}
  \includegraphics[width=0.95\linewidth]{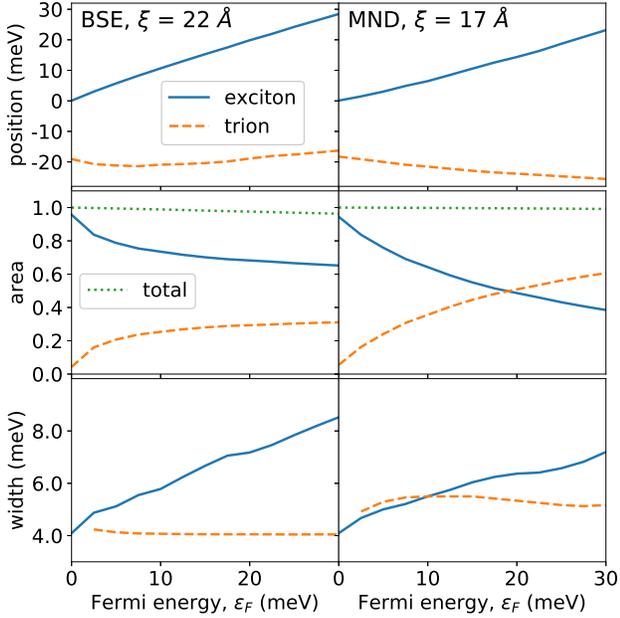}
  \caption{The doping-dependent exciton-trion peak positions, peak areas, and peak widths
  calculated by the BSE formalism (left) and the MND theory (right) with
  $m_{\text{e}}=0.045\;(\text{eV})^{-1}\text{\AA}^{-2}$, $\beta^{-1}=0.1$ meV, $\gamma=2$
  meV, $N=140$. The peak areas are calculated by integrating the imaginary part of the
  exciton Green's function, and are normalized to the total area at
  $\varepsilon_{\text{F}}=0$ for each method.}
  \label{fig:peak}
\end{figure}

By altering the Fermi energy, the doping dependence of the optical spectrum can be
studied. Fig.~\ref{fig:doping_dependent_spectra} shows the doping-dependent optical
spectra calculated by both the BSE formalism and the MND theory. As can be seen, for both
methods the trion peak emerges with positive Fermi energies, and the energy splitting
between the exciton peak and the trion peak increases upon doping. The oscillator strength
transfer from exciton peak to trion peak calculated by the BSE formalism saturates with
increasing $\varepsilon_{\text{F}}$. The saturation can only be attributed partially to
the Pauli-blocking effect discussed in Sec.~\ref{sub:oscillator}. Via performing the BSE
calculation without the distribution function of allowed transition $\rho_{n,\mathbf{Q}}$
for $n=0$ (the trion bound state) in Eq.~(\ref{BSE_self_energy}), a new doping-dependent
spectra is shown in the inset of Fig.~\ref{fig:doping_dependent_spectra} and a reduction
of the saturation of the oscillator strength transfer is observed. The heights of the
trion peaks in the high doping-density regime
($\varepsilon_{\text{F}}\geq\Delta_{\text{T}}$) surpass those of the exciton peaks.
However, the peak area transfer between the exciton and trion peaks remains saturated. The
result (the exceedance of the peak heights and the saturation of the peak areas) is
consistent with the calculation of Ref\cite{efimkin2017many}, which also does not appear
to contain explicit Pauli-blocking term for the trion transition. On the other hand, the
oscillator strength transfer calculated by the MND theory shows no saturation.

Via simple curve fitting, the energy splitting between exciton peak and trion peak, the
peak areas, and the peak widths of the two peaks calculated by the BSE formalism and the
MND theory are estimated and shown in Fig.~\ref{fig:peak}. For calculations based on the
BSE formalism, the energy splitting is approximately proportional to the Fermi energy,
$E_{\text{split}}\simeq\Delta_{\text{T}}+\varepsilon_{\text{F}}$, and $\Delta_{\text{T}}$
is about $19$ meV, where the trion binding energy is equal to the peak position of the
Diff.\,DOS calculation shown in Fig.~\ref{fig:DOS_spectra} (a) with $\xi=22\;\text{\AA}$
and $m=0.030\;\text{eV}^{-1}\text{\AA}^{-2}$. The increasing peak area of the trion
transition and the decreasing peak area of the exciton transition with respect to
increasing the Fermi energy are shown, and the total area is conserved for different Fermi
energies. As can be seen in Fig.~\ref{fig:peak}, the oscillator strength transfer is
gradually saturated. For the peak widths, the exciton linewidth is roughly proportional to
the Fermi energy, implying that the scattering potential is more similar to a contact
potential at the relevant length scales. On the other hand, the trion linewidth is
basically invariant to changes in the Fermi energy as found in Ref\cite{efimkin2017many}.

For the calculations based on the MND theory, the trion peak continues to grow as the
Fermi energy increases, and the energy splitting is given by
$E_{\text{split}}\simeq\Delta_{\text{T}}+\varepsilon_{\text{F}}$, with $\Delta_{\text{T}}$
about $19$ meV, where the trion binding energy is equal to the peak position of the
Diff.\,DOS calculated in Fig.~\ref{fig:DOS_spectra} (a) with $\xi=17\;\text{\AA}$ and
$m=0.045\;\text{eV}^{-1}\text{\AA}^{-2}$. Note that the doping density is nonzero at
$\varepsilon_{\text{F}}=0$ eV due to the small but non-zero temperature, such that the
trion lineshape emerges and has a finite oscillator strength even at
$\varepsilon_{\text{F}}=0$. Similar to the BSE calculation, with increasing Fermi energy
the peak area of the trion transition increases and the peak area of the exciton
transition decreases, with the total area conserved for different doping densities. The
peak areas of exciton and trion transitions intersect around $18$ meV, which is also close
to the trion binding energy. The peak widths of exciton and trion transitions are found to
have a somewhat quantitatively different dependence with the Fermi energy. The trion
linewidth calculated by the MND theory is roughly invariant to the Fermi energy, and the
exciton linewidth is proportional to the Fermi energy, as in the BSE calculation.

In comparison with the results calculated by the BSE formalism, the MND results show
better correspondence with at least some experiments which show an oscillator strength
transfer that shows no sign of saturation with increasing
$\varepsilon_{\text{F}}$\cite{kheng1993observation, huard2000bound, yang2015robust}. This
implies that the Fermi sea multiple-electron-hole excitations, which are responsible for
the creation of the Fermi-edge singularity, may also contribute to the trion transition.

\subsection{Temperature-dependent optical spectra\label{sub:temperature}}

\begin{figure}
  \includegraphics[width=0.95\linewidth]{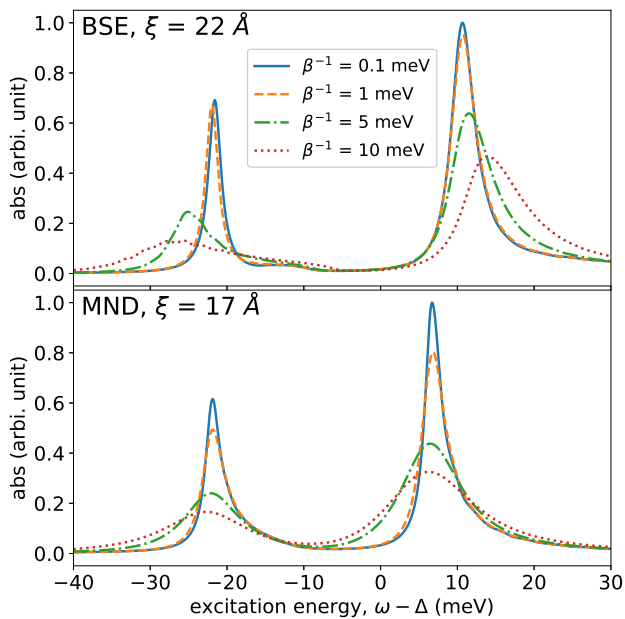}
  \caption{The temperature-dependent optical spectra calculated by the BSE formalism
  (upper panel) and the MND theory (lower panel) with
  $m_{\text{e}}=0.045\;(\text{eV})^{-1}\text{\AA}^{-2}$, $\varepsilon_{\text{F}}=10$ meV,
  $\gamma=1$ meV, $\varepsilon_{\Lambda}=2.0$ eV, $N=120$.}
  \label{fig:temp_spectra}
\end{figure}

In this section, the temperature dependence of the optical spectrum is discussed. In
Fig.~\ref{fig:temp_spectra}, the temperature-dependent optical spectra calculated by both
the BSE formalism and the MND theory are given. For both spectra, the linewidths are
broadened and the peak-heights become lower as the temperature increases. For the spectra
calculated by the MND theory in extremely low temperature regime (below $1$ meV), both the
exciton and the trion peaks exhibit asymmetric lineshapes near the transition energies
($\omega-\Delta\simeq\mu$ for exciton and $\omega-\Delta\simeq-\Delta_{\text{T}}$ for
trion), and the peak heights are sensitive to the temperature. Both of these features are
signatures of the Fermi-edge singularity. On the other hand, the spectra calculated by the
BSE formalism are relatively insensitive to temperature variations in this low temperature
regime. As can be seen in Fig.~\ref{fig:temp_spectra}, only very small variations of the
trion and exciton lineshapes can be observed when the temperature is on the order of 5\%
of the trion binding energy.

\section{Discussions and conclusion\label{sec:conclusion}}

In the present work, we have theoretically studied the problem of an exciton immersed in a
Fermi sea which interacts with electrons through a scattering potential. We have focused
on two approximate methods, the BSE formalism and the MND theory, to solve a many-body
Hamiltonian parametrized to describe two-dimensional semiconductors. We find some results
that are coincident with experimental observations and expectations. Both the BSE
formalism and the MND theory describe the trion peak emergence, the oscillator strength
transfer, and the doping-independent lineshapes in a sensible manner in the low
doping-density regime. When the Fermi energy exceeds the trion binding energy, the BSE
formalism may not account completely for the oscillator strength transfer from the exciton
peak to the trion peak, while the MND theory is found to describe this effect. This
limitation of the BSE approach can only be altered in part by the \textit{ad-hoc}
procedure of removing the Pauli-blocking term.

Neither the BSE nor the MND theories are expected to capture all experimental features
over all parameter regimes, since they only describe two limiting physical situations of a
simple model. Even if we could solve the many-body scattering Hamiltonian exactly, there
are still many factors which are not included in our model which could affect the results
of the calculations. In Sec.~\ref{sub:DOS} we have discussed three such factors that may
result in an underestimation of the trion binding energy. The same missing physical
ingredients may also cause errors in the prediction of the doping dependence and
temperature dependence of optical spectra. In addition to the mentioned factors, a
significant approximation of the many-body scattering Hamiltonian is that the Fermi sea is
assumed to be composed of non-interacting electrons. This is not a particularly realistic
assumption, since in two dimensions the long-range Coulomb interaction is not fully
screened and thus electron correlation effects may be important. A direct consequence of
electron correlation is that the electron mass, the exciton mass, and the scattering
potential should be renormalized according to the Fermi energy and
temperature\cite{mahan2013many, galitski2004universal, zhang2005quasiparticle}. Electron
correlation can also affect the ratio between the electron mass and exciton mass, and
alter the effective scattering potential, thus changing the trion binding energy and the
absorption lineshapes. Without considering the renormalization induced by electron
correlation, precise predictions are difficult to obtain. These issues are beyond the
scope of the present work, and we leave the topic of electron-electron interactions in the
optical spectra to a future study.

Despite the physical factors which we do not include, the present model and its
approximate solutions still provide a qualitative description and a physically clear
picture of the doping-dependent optical spectra in two-dimensional semiconductors. In a
nutshell, trion formation can be realized as the dynamical generation of a trion and a
hole in the Fermi sea. The dynamical process originates from the electron-hole
polarization near the Fermi sea induced by electron-exciton interaction, with the trion
state formed as the bound state of an electron-exciton scattering process. From the
fundamental (electron/hole) particle point of view, the scattering event involves a
four-particle generation process, with the participation of two electrons plus one hole in
the conduction band, and one hole in the valence band. The wavefunction for the trion-hole
state is coherently coupled with the exciton wavefunction, as the discussion in
Sec.~\ref{sec:wavefunction}. The coupling strength can be connected to the
electron-exciton scattering potential. Recently, two-dimensional coherent spectra in
quantum wells\cite{moody2014coherent} and TMDCs\cite{singh2014coherent, hao2016coherent,
shepard2017trion} have been studied, and the role of coherent exciton-trion coupling has
been discussed in the context of these experiments. Coherent coupling has been attributed
to exciton-trion many-body interactions\cite{singh2014coherent}, but a microscopic theory
of these interactions has not been put forward. The present theory supports the existence
of such a coupling, and may provide a theoretical foundation for the study coherent
multi-dimensional spectra in the future.

{\em Note added} - After this work was completed, we became aware of
Ref.\cite{chang2018crossover} which presents calculations related to, and several
conclusions similar to, those found in this work.

\section*{acknowledgements}

One of authors (Y.-W.C.) wants to thank the financial support from the Postdoctoral
Research Abroad Program of Ministry of Science and Technology, Taiwan, Republic of China.
D.R.R. was supported by NSF-CHE 1839464. We wish to thank Alexey Chernikov, Misha Glazov
and Archana Raja for discussions and collaboration connected to this work.

\appendix

% \input{appendix}

% \section{Reference}

% \input{reference}

\end{document}